\documentclass[twocolumn]{aastex631}

\shorttitle{Exploring magnetic loops and serpentine fields in the quiet Sun with the GRIS-IFU}
\shortauthors{Campbell et al.}

\begin{document}

\title{Exploring magnetic loops and serpentine fields in the quiet Sun with the GRIS-IFU}

\author[0000-0001-5699-2991]{Ryan J. Campbell}
\affiliation{Astrophysics Research Centre,
Queen's University of Belfast,
Northern Ireland, BT7 1NN, UK}

\author[0000-0003-4920-0153]{Ricardo Gafeira}
\affiliation{Instituto de Astrof\'isica de Andalucía, Apartado de Correos 3004, 18080 Granada, Spain}
\affiliation{Instituto de Astrofísica e Ciências do Espaço, Departamento de Física da Universidade de Coimbra,\\ Rua do Observatório s/n, 3040-004 Coimbra, Portuga}

\author[0000-0002-7725-6296]{Mihalis Mathioudakis}
\affiliation{Astrophysics Research Centre,
Queen's University of Belfast,
Northern Ireland, BT7 1NN, UK}

\author[0000-0001-5518-8782]{Carlos Quintero Noda}
\affiliation{Instituto de Astrof\'isica de Canarias,
V\'ia L\'actea s/n, E-38205 La Laguna,
Tenerife, Spain}
\affiliation{Departamento de Astrof\'isica, Universidad de La Laguna, E-38206 La Laguna, Tenerife, Spain}

\author[0000-0002-6210-9648]{Manuel Collados}
\affiliation{Instituto de Astrof\'isica de Canarias,
V\'ia L\'actea s/n, E-38205 La Laguna,
Tenerife, Spain}
\affiliation{Departamento de Astrof\'isica, Universidad de La Laguna, E-38206 La Laguna, Tenerife, Spain}

\begin{abstract}
{Synthetic observations produced from radiative magnetohydronamic simulations have predicted that higher polarization fractions in the quiet solar photosphere would be revealed by increasing the total integration time of observations at GREGOR resolutions.} 
{We present recently acquired disk centre observations of the Fe I $15648.5$ $\mathrm{\AA}$ line obtained with the GREGOR telescope equipped with the GRIS-IFU during excellent seeing conditions, showing exceptionally high polarization fractions.}
{Our observation reveal an internetwork region with a majority ($>60\%$) of magnetised pixels displaying a clear transverse component of the magnetic field. This result is in stark contrast to previous disk-centre GRIS-IFU observations in this spectral line, which had predominantly vertical magnetic fields in the deep photosphere. At the same time, the median magnetic field strength is weaker than previous GRIS-IFU observations, indicating that the larger fraction of polarization signals cannot be explained by a more active target. We use the Stokes Inversion based on Response functions (SIR) code to analyse the data, {performing over $45$ million inversions}, and interrogate the impact of two conflicting approaches to the treatment of noise on the retrieval of the magnetic inclination and azimuth. We present several case studies of the zoo of magnetic features present in these data, including small-scale magnetic loops that seem to be embedded in a sea of magnetism, and serpentine fields, focusing on regions where full-vector spectropolarimetry has been achieved. We also present a new open-source Python 3 analysis tool, SIR Explorer (SIRE), that we use to examine the dynamics of these small-scale magnetic features.}
\end{abstract}


\section{Introduction} \label{sec:intro}
\cite{lites2008} revealed the quiet Sun (QS) internetwork (IN) as dominated by horizontal (i.e. transverse with respect to the solar normal) magnetic fields at an effective spatial resolution of $0.3''$. However, this result is not undisputed and remains subject to contradiction by other studies, as reviewed by \cite{steiner2012}. For instance, the lack of variation in the degree of linear and circular polarisation recorded in near infrared (NIR) observations by \cite{marian2008_mu} at $0.8''$ as a function of different heliocentric angles points to a QS magnetic field which has no preferential bias in orientation. If the distribution of magnetic field inclinations is isotropic, its probability density function is given as,
\begin{equation}
    P(\gamma) = \frac{\sin \gamma}{2},
\end{equation}
where $\gamma$ is the magnetic inclination angle, {defined as the angle between the magnetic vector and solar normal}, such that $P(\gamma)$ has a maximum at $90^\circ$. It is perhaps counter-intuitive, but this would mean most of the fields are transverse, because for the magnetic field to be aligned along the line-of-sight (LOS) it has to point in one of two possible directions, but to be transverse there are many more possible directions \citep{almeida2011}.

\begin{figure*}
\centering
\includegraphics[width=.9\textwidth]{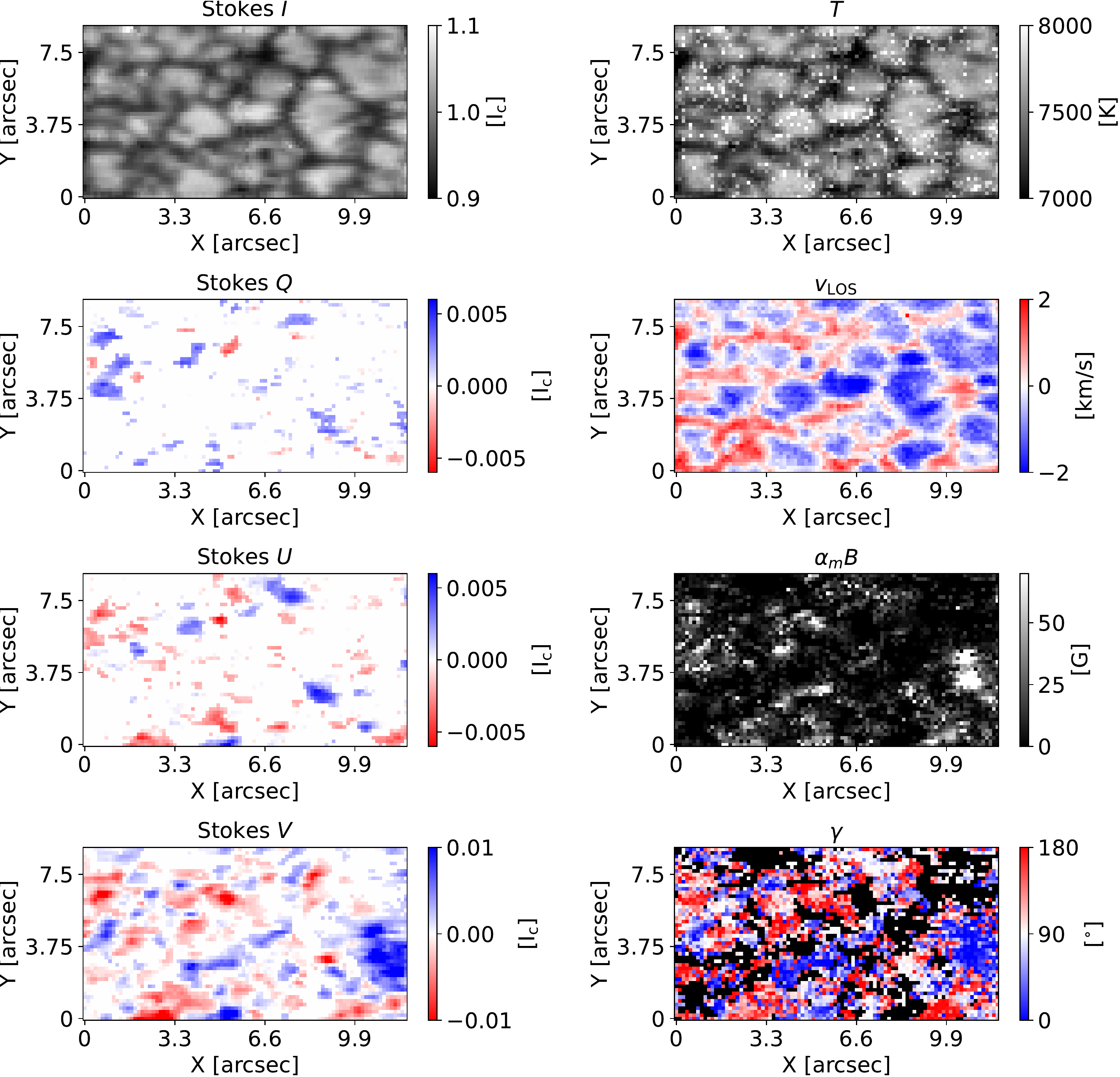}
\caption{Sample frame from a GRIS-IFU scan of a quiet Sun region on the $24$ August $2021$ (scan D) {showing in the left column from top to bottom the continuum-normalized Stokes $I$, $Q$, $U$, and $V$, and in the right column from top to bottom the $T$, $v_{\mathrm{LOS}}$, $\alpha_m B$, and $\gamma$ as derived from SIR inversions (scenario 3)}. Stokes $I$ is shown at a wavelength of $15650.59$ $\mathrm{\AA}$ in the continuum and the polarization profiles are shown at $15648.36$ $\mathrm{\AA}$ in the blue lobes of the $g_{\mathrm{eff}} = 3$ line. Any pixel which does not have a Stokes $Q$, $U$, or $V$ profile with maximum unsigned amplitude across the $15648.5$ $\mathrm{\AA}$ line below the $5\sigma_n$ threshold has been set to zero and masked from the plots of Stokes $Q$, Stokes $U$, Stokes $V$, $\alpha_m B$, and $\gamma$. $T$ is shown at $log\tau_{5000\mathrm{\AA}} = 0.5$, while the other model parameters are shown at constant in depth. }
          \label{fig:map}%
\end{figure*}

The greatest difficulty in constraining the $\gamma$ from inversions of IN observations in a consistent way results from the differing treatments of varying levels of noise. In the weak field regime, a vertical field produces a larger amplitude circular polarization profile than the amplitude of a linear polarization profile produced by a horizontal field of equal strength at disk centre. This creates an intrinsic bias against being able to confidently detect horizontal fields at disk centre when a given noise threshold is equally applied to Stokes $Q$, $U$, and $V$. By producing models with deliberately purely vertical fields, synthesizing the Stokes vector, and adding noise before inverting again, \cite{Borrero2011} show that an inversion code will return an overabundance of horizontal inclinations. {As a result, even when one inverts only those pixels with at least one Stokes $Q$, $U$, or $V$ profile above a noise threshold, and most of the pixels have only Stokes $V$ above the threshold, this arguably results in a possible bias in favour of horizontal fields as the inversion code interprets noise in Stokes $Q$ and $U$ as real signals.}

The deepest Hinode/SP integrations show circular and linear polarisation in $88\%$ and $53\%$ of the field of view (FOV), respectively, but this comes at the expense of spatio-temporal resolution which distorts the polarization signals \citep{rubio2012}. Most recently, observations of the IN with visible lines at the ground-based Swedish Solar Telescope (SST; \cite{gosic2021,gosic2022,ledvina2022}) and balloon-borne Sunrise experiment \citep{danilovic_imax_2010,marian2012,kianfar2018} have provided good statistics in terms of the longitudinal field and even of cancellations, but visible photospheric lines still struggle to confidently detect the horizontal fields that should be present in magnetic loops along the polarity inversion line (PIL) without significant spatial, spectral, or temporal binning. Observations with the NIR line pair at GREGOR with the GREGOR Infrared Spectrograph (GRIS; \cite{lagg2016,marian2016}) have demonstrated higher efficacy at measuring linear polarization at similar spatial resolutions. This is despite the fact that modelling with simulations by \cite{steiner2008} shows the vertical field being prominent in the deep photosphere, where the NIR line pair is sensitive \citep{quintero2021}, with the horizontal component of the magnetic field becoming increasingly important in the upper photosphere, where the visible line pair is sensitive, with convective overshooting responsible for expelling horizontal fields to greater heights. \cite{Campbell2021a} have used the Integral Field Unit (IFU) mode of GRIS to examine the dynamics of small-scale magnetic features, including magnetic loops. Two-dimensional spectrographs like the GRIS-IFU are uniquely capable of providing time-series imaging with high polarimetric sensitivity and spectral resolution simultaneously, with the size of the FOV and cadence acting as competing factors. 

\cite{Campbell2021b} have predicted, using synthetic observations produced from MURaM simulations \citep{vogler2005} degraded to GREGOR/GRIS-IFU resolutions, that increasing the total integration time of the observations will yield higher fractions of polarization and allow the magnetic inclination to be better constrained in a larger fraction of the FOV. We report observations calibrated based on these predictions.

\section{Observations}\label{sect:observations}

Repeated observations of disk centre QS regions were taken in late August 2021 using the GRIS-IFU \citep{DominguezTagle2022} at GREGOR \citep{Schmidt2012, kleint2020}. The GRIS-IFU was operated in double sampling mode, with a $3$ (vertical) by $2$ (horizontal) mosaic pattern, with an exposure time of $60$ ms per polarimetric state, and 20 accumulations (for a total integration time per pixel of $4.8$ seconds). This resulted in a cadence of $103$ seconds between frames. Compared to similar $2019$ scans, the exposure time and accumulations were increased to target a higher S/N, but the number of steps in the mosaic pattern was reduced to constrain the cadence between frames. The FOV of the scans is therefore $12.15''$ by $9.024''$ with a spatial sampling of $0.135''$ by $0.188''$ in the $x$- and $y-$directions, respectively.

The GRIS-IFU observed a 40 Å spectral window that includes the highly magnetically sensitive photospheric Fe I line at $15648.5$ $\mathrm{\AA}$. The spectral dispersion was determined to be $39.83$ m$\mathrm{\AA}$/pixel by comparison with a degraded Fourier Transform Spectrometer (FTS) atlas \citep{atlas}.

During the observing campaign we benefitted from very good seeing conditions. Table \ref{table:raw_stats} lists five datasets which are candidates for analysis as they were taken with good to excellent seeing, as quantified by the locally measured Fried parameter, $r_0$, and the root-mean-square continuum intensity contrast, $\delta I_{\mathrm{rms}}$. The GRIS reduction pipeline \citep{gregor2012} was employed for dark current removal, flat fielding, polarimetric calibration, and cross-talk corrections. 

\section{Results}\label{sect:results}
\subsection{Polarization analysis}\label{sect:polarization}
In order to quantify the fraction of pixels exhibiting a polarization profile confidently above the noise level in the datasets listed in Table \ref{table:raw_stats}, we adopt the  method used by \cite{lagg2016,Campbell2021a}. The purpose of this analysis is to enable a comparison between the new GRIS-IFU datasets, previous GRIS-IFU datasets, and datasets from other facilities. For a given Stokes parameter, we determine the noise level, $\sigma_n$, by calculating the standard deviation in the continuum {in each frame}. The explicit assumption is made that the continuum is unpolarized and the primary type of noise is photon noise. We then determine whether a pixel has a Stokes $Q$, $U$, or $V$ profile with maximum amplitude greater than a threshold set at $5\sigma_n$ across the $15648.5$ $\mathrm{\AA}$ line. If a given pixel has a maximum amplitude in Stokes $V$ greater than the threshold, the pixel is said to have a confidently measured circular polarization (CP) signal. Similarly, if a given pixel has a maximum amplitude in Stokes $Q$ or $U$ greater than the threshold, the pixel is said to have a confidently measured linear polarization (LP) signal. If a Stokes vector has neither LP or CP, it is said to have no polarization (NP).
 
Table \ref{table:raw_stats} shows the time-averaged results of this analysis. The exceptionally high polarization fractions of scans D and E stand out as being much larger than scans A, B, and C. Figure \ref{fig:map} shows a sample frame from scan D. This frame had a $\delta I_{\mathrm{rms}}$ of $3.4\%$, a maximal $r_0$ of $22$ cm, and shows an abundance of polarization. In terms of understanding why the polarization fractions of scan D and E are higher than the other scans, the first factor to consider is that the GRIS-IFU FOV is very small. There is therefore always a risk when observing with the GRIS-IFU that one could point to a so-called `void', where there is little detectable magnetic flux, or indeed the opposite. The second factor to consider is the quality and stability of the seeing conditions. From the $r_0$ and $\delta I_{\mathrm{rms}}$ values for each of the scans, it is clear that the seeing conditions for scans D and E were superior in both quality and stability. For instance, scan A had a peak $\delta I_{\mathrm{rms}}$ of only $3\%$ and a low of $1.8\%$, while scan D had a peak of $3.4\%$ and low of $2.7\%$. This is important as atmospheric stability is essential to achieve a maximal effective spatial resolution, and impacts heavily on the polarization fractions recorded because opposite polarity profiles cancel out within the spatio-temporal resolution element with insufficient spatial resolution. 
\begin{table}
\caption{Time-averaged percentage of linear (LP) and circular (CP) polarization profiles with maximum amplitudes above $5\sigma_n$ for level 1 GRIS-IFU/GREGOR data. The percentages are calculated relative to the full FOV. The $1\sigma_n$ noise level is determined by the calculation of the standard deviation in the relevant Stokes parameter at continuum wavelengths in each frame. All scans were taken in August 2021 and the times are given in universal time.}       
\label{table:raw_stats}      
\centering                          
\begin{tabular}{|c| c | c c c c |}        
\hline               
 Label & Day, time & $\%$LP & $\%$CP & $\%$LP $\&$ CP & $\%$NP \\
\hline      
    A& 19, 08:14 & 12.7 & 43.4 & 8.6 & 52.5  \\ 
    B& 20, 08:56  & 12.4 & 42.2 & 7.4 & 52.8  \\ 
    C& 23, 08:00  & 6.7 & 28.7 & 3.7 & 68.2  \\ 
    D& 24, 07:50  & 29.3 & 64.7 & 22.7 & 28.6  \\ 
    E& 24, 08:55  & 30.7 & 66.5 & 24.6 & 27.4 \\ 
\hline      
\end{tabular}
\end{table}

\begin{table}
    \caption{As for Table \ref{table:raw_stats}, but for the PCA-RVM reconstructed GRIS-IFU data. The noise threshold is set at 5$\sigma_n$ as measured in the continuum of the original level 1 data.}       
    \label{table:5sig_PCARVM_stats}      
    \centering                          
    \begin{tabular}{|c|cccc|}        
    \hline
      Label & $\%$LP & $\%$CP & $\%$LP $\&$ CP & $\%$NP  \\   
    \hline 
      D & 22.0 & 59.0 & 16.2 & 35.3 \\
    \hline 
      E & 23.2 & 60.7 & 17.6 & 33.8 \\
      
    \hline
    \end{tabular}
\end{table}

\begin{figure}
   \centering
   \includegraphics[width=.9\columnwidth]{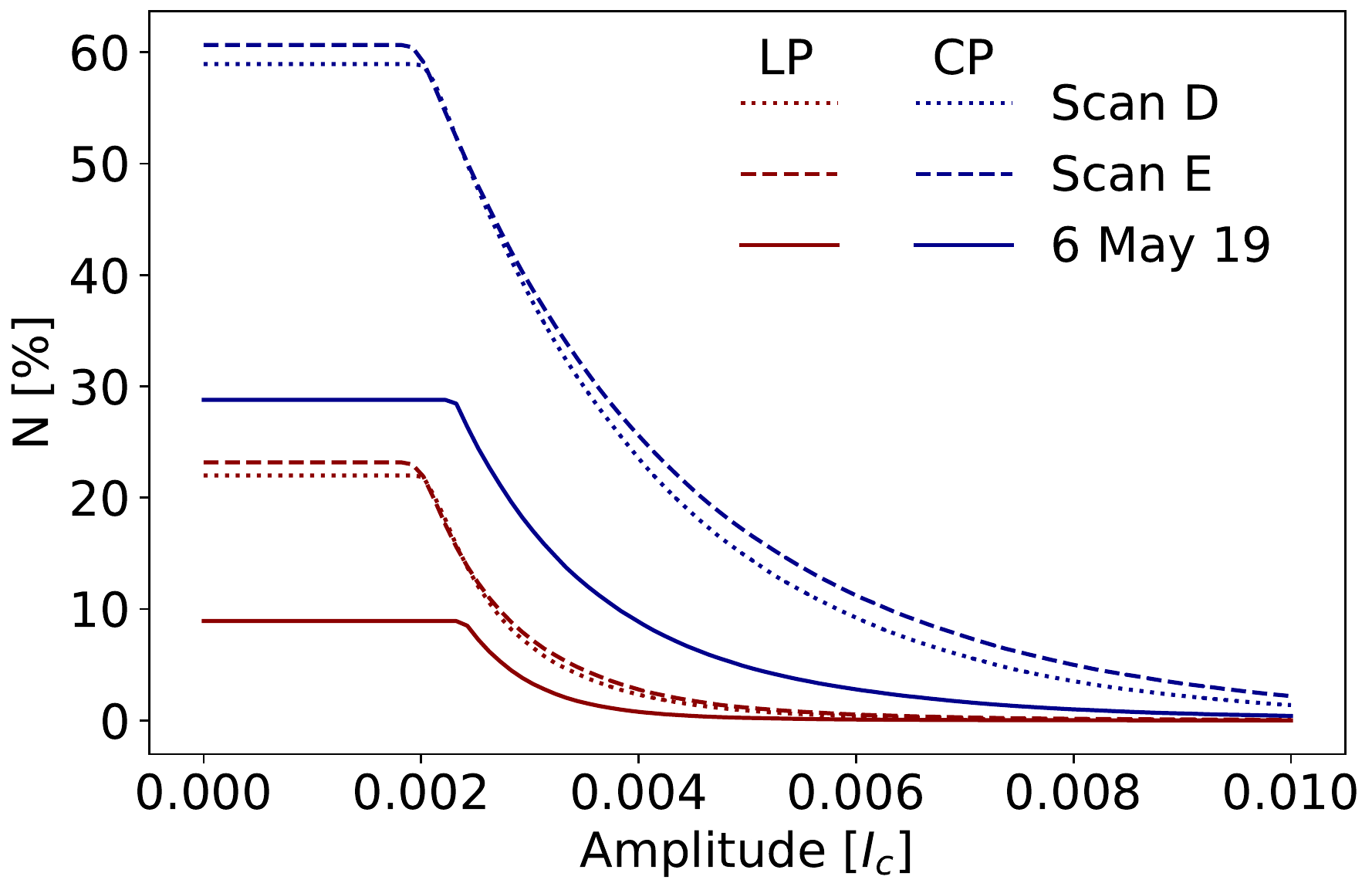}

   \caption{Area occupied by pixels with a LP (\textit{red lines}) or CP (\textit{blue lines}) signal for scan D (\textit{dotted lines}), scan E (\textit{dashed lines}), and the scan taken on the $6$ May 2019 (\textit{solid lines}) {for given amplitudes}. {The data has undergone PCA-RVM reconstruction and had any Stokes profiles with maximum unsigned amplitude across the $15648.5$ $\mathrm{\AA}$ line below the $5\sigma_n$ threshold in the 2021 data, and $3\sigma_n$ in the 2019 data, set to zero.}}
              \label{fig:statsPCARVM}%
    \end{figure}

Principal component analysis (PCA) with $15$ retained eigenvectors is applied to scans D and E for the purpose of removing noise. As eludicated by \cite{Borrero2011}, while the probability of photon noise producing a signal greater than $3\sigma_n$ is only $0.3\%$, the probability increases dramatically when more and more wavelength points are considered. By applying PCA to remove noise and using a stringent $5\sigma_n$ noise threshold, we are reducing the probability that a false signal will survive and enter the analysis. As interference fringes are present {in the wavelength domain} of the polarization profiles in some pixels, a relevance vector machine (RVM) is employed to remove these defects and produce apparently noiseless polarization profiles. The full reconstruction process is the same as employed by \cite{Campbell2021a} on GRIS-IFU/GREGOR data and is described in detail therein. By applying the PCA-RVM reconstruction process to scans D and E, this enables a comparison with the PCA-RVM reconstructed statistics reported by \cite{Campbell2021a}. For this purpose in this paper we will use the scan taken on $6$ May 2019, henceforth referenced simply as the 2019 data.

Table \ref{table:5sig_PCARVM_stats} shows the time-averaged polarization fractions for scan D and E {after application of the PCA-RVM reconstruction process}. Figure \ref{fig:statsPCARVM} shows the time-averaged polarization fractions of scans D and E as a function of amplitude alongside the scan taken on $6$ May $2019$. In this case, any Stokes profile whose maximum unsigned amplitude across the $15648.5$ $\mathrm{\AA}$ line does not exceed the $5\sigma_n$ threshold for the two 2021 datasets, and $3\sigma_n$ for the 2019 dataset, has been set to zero. From Fig.~\ref{fig:statsPCARVM} it is clear that the $5\sigma_n$ threshold is on average {at a slightly lower amplitude} in scans D and E than the $3\sigma_n$ threshold in the 2019 scan. Seeing-induced cross-talk becomes more likely with longer exposures, as with the longer total integration time it is more likely that the modulations of the Stokes parameters will not be conducted fast enough for Earth's atmosphere to be considered `frozen' during measurements \citep{crosstalk1999}. We choose a stringent $5\sigma_n$ threshold for the August 2021 data {that still allows us to access weaker polarization signals} than in the 2019 data, {because the former has a lower noise level than the latter}, but is sufficiently large to {allow us to assume any seeing-induced cross-talk produced is unlikely to have an amplitude of this magnitud}e. There is a significantly higher fraction of CP and LP in the new scans than the old scans. Perhaps the most significant difference between the old and new scans is the much higher percentage of the FOV which has both LP and CP. Previously, between $3-5\%$ of the FOV had both LP and CP, but in the new scans this number has significantly increased to $16-18\%$ at $5\sigma_n$. This means that $\gamma$ should be retrievable from inversions in a much higher fraction of the FOV and it would also be anticipated that a larger fraction of pixels would harbour fields with an intermediate (as opposed to highly vertical or highly inclined) inclination.

      

\subsection{Inversion strategies}\label{sect:inversions}
We use the Stokes Inversion based on Response functions (SIR) \citep{SIR} code to invert scans D and E. We adapt the parallelized wrapper to SIR written by \cite{ricardo2021} to enable the kinematic and magnetic parameters of the input model(s) to be randomized within physically sensible upper and lower boundaries at every iteration. This is an essential step to maximise the statistical chances of achieving the global $\chi^2$ minimum solution for each Stokes vector. For the inversion, we adopt the same scheme as tested on real observations of the NIR line pair by \cite{marian2016, Campbell2021a} and on synthetic observations by \cite{Campbell2021b}. This is a two-component inversion with one non-magnetic (i.e. the magnetic field strength, $B$, was set to zero and was not allowed to vary) and one magnetic model, which are combined by SIR in accordance to their respective filling factors, $\alpha$, which is another free parameter. For clarity, we refer to the filling factor of the magnetic model as $\alpha_m$. {SIR allows the user to determine the optical depths at which perturbations will be made between iterations by selecting a number of nodes in each variable parameter, with the full parameter stratification in optical depth given by cubic splines or linear interpolation. }Apart from temperature, $T$, for which $4$ nodes were selected, each parameter, {including $B$}, was forced to be constant in optical depth. The macroturbulent velocity, $v_{\mathrm{mac}}$, was included as a free parameter in the non-magnetic model and forced to be the same in the magnetic model. The $T$ was also determined by the non-magnetic model and forced to be the same in the magnetic model. The line of sight velocity, $v_{\mathrm{LOS}}$, and microturbulent velocity, $v_{\mathrm{mic}}$, were free parameters in both models. Full PCA-RVM reconstruction was applied to the datasets before inversion. We used the empirically determined atomic parameters made available by \cite{lines2021} and abundances from \cite{asplund}. Included in the inversion were seven spectral lines, including five Fe I lines with rest wavelengths of $15645.02$ $\mathrm{\AA}$, $15645.303$ $\mathrm{\AA}$, $15648.514$ $\mathrm{\AA}$, $15652.873$ $\mathrm{\AA}$, $15662.017$ $\mathrm{\AA}$, $15665.241$ $\mathrm{\AA}$, one blended Fe II line with rest wavelength $15648.514$ $\mathrm{\AA}$, and one blended O I line with rest wavelength $15665.098$ $\mathrm{\AA}$.

 

  
  


  
We interrogate the impact of noise on the retrieval of $B$, $\gamma$, and $\alpha_m$ statistically by taking more than one approach to data treatment. In order to accurately determine $\gamma$, one must measure both linear polarization and circular polarization confidently above the noise level. However, it can be argued that the noise level itself places limits on the amplitude of the weakest polarization profiles. We divide the approaches into three scenarios. In the first scenario (S1), the full datasets were inverted with all Stokes parameters provided to SIR for each pixel without any consideration for signal amplitudes relative to the $\sigma_n$ level. In scenario 2 (S2), any individual Stokes $Q$, $U$, or $V$ profile in a given pixel whose maximum amplitude across the $15648.5$ $\mathrm{\AA}$ line does not reach the $5\sigma_n$ threshold was set to zero before inversion. In scenario 3 (S3), any individual Stokes $Q$, $U$, or $V$ profile that does not reach the $5\sigma_n$ threshold was set to zero before inversion as in S2, {with the exception that} if one linear polarization parameter was confidently measured then the weaker linear polarization parameter was not set to zero {despite being below the threshold}. This third scenario is included to investigate whether only one linear polarization parameter (i.e. cases where Stokes $Q$ is $>5\sigma_n$ but Stokes $U$ is not, or vice versa) is sufficient to constrain $\gamma$ and if there are any impacts on other magnetic parameters when the weaker linear polarization parameter is set to zero and the stronger one is not. Each inversion was repeated $50$ times with randomized initial models, resulting in over $45$ million inversions in total for all three scenarios and both datasets. 

\begin{figure*}
\centering
\includegraphics[width=.79\columnwidth]{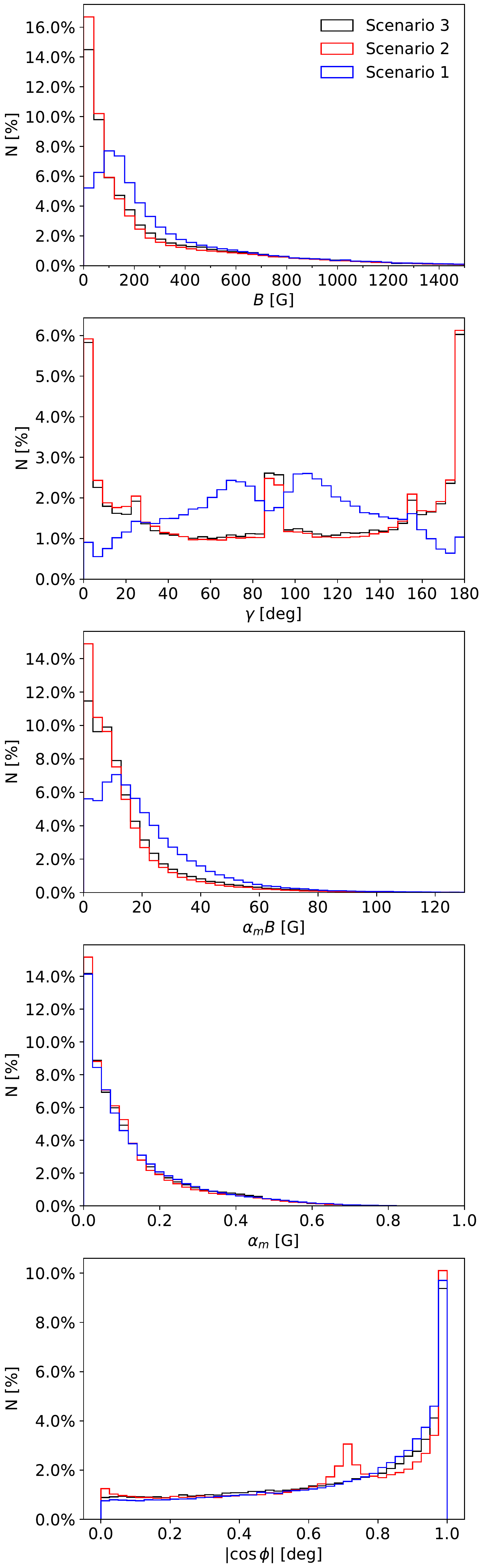}
\includegraphics[width=.79\columnwidth]{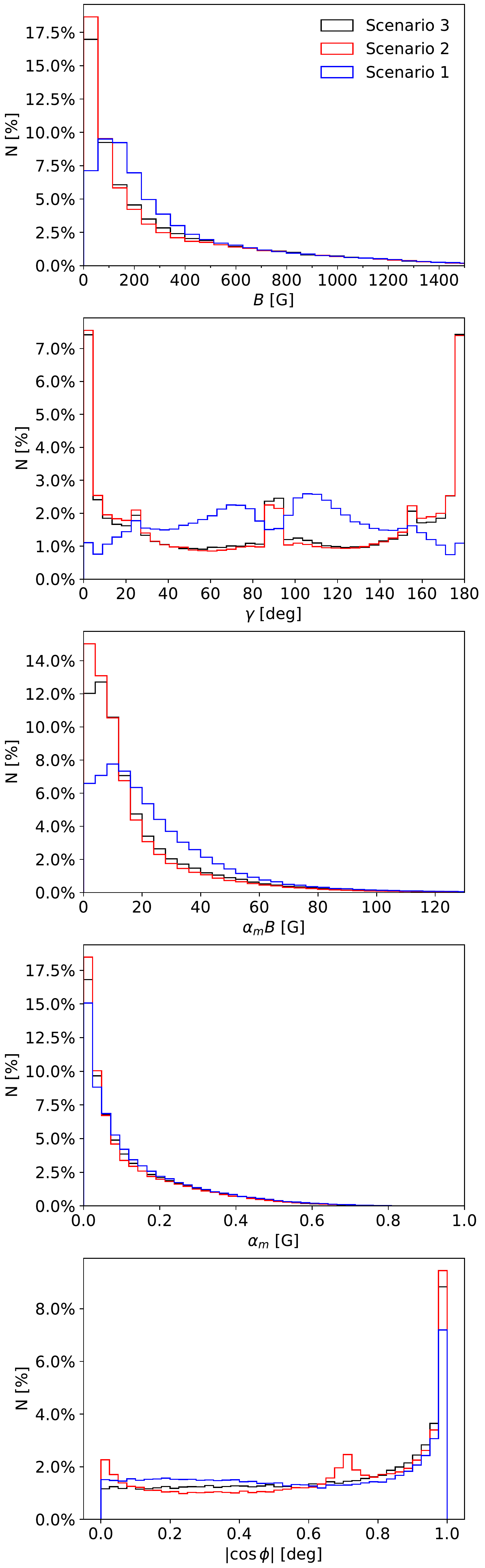}
\caption{Histograms of $B$ (\textit{top row}), $\gamma$ (\textit{second row}), $\alpha_m B$ (\textit{third row}), $\alpha_m$ (\textit{fourth row}), and $\left|\cos\phi\right|$ (\textit{bottom row}) returned from the inversions. The histograms are shown for three inversion scenarios {(S1, S2, and S3)} as described in the text. The \textit{left} and \textit{right} columns show the distributions for scan D and E, respectively.  Histograms are weighted with respect to the total number of pixels, i.e. including those with no polarization. Pixels which had no measured polarization in any Stokes parameter are excluded from the histogram (i.e. only pixels with maximum absolute amplitude across the $15648.5$ $\mathrm{\AA}$ line greater than the $5\sigma_n$ threshold in at least one Stokes parameter are included).}
          \label{fig:inv_parameters}%
\end{figure*}

\subsection{Magnetic filling factors and field strengths}
Figure \ref{fig:inv_parameters} shows the distributions of $B$, $\gamma$, $\alpha_m B$, $\alpha_m$, and $\left|\cos\phi\right|$, where $\phi$ is the azimuthal angle of the magnetic field vector in the plane perpendicular to the observer's LOS, for each scenario and scans D and E. First, the $\alpha_m$ distribution looks very similar in all scenarios. However, the $B$ distribution is significantly different between S1 on the one hand and S2 and S3 on the other hand below $600$ G for both scans. In S1, the $B$ distribution peaks at around $100$ G, but for S2 and S3 there is a much larger number of pixels with $B$ values below $100$ G. {Since $B$ is determined by the longitudinal and transverse components, setting linear and circular polarization signals to zero naturally results in lower values of $B$ overall.} As a consequence, the $\alpha_m B$ distributions for S1 on one hand and S2 and S3 on the other hand also differ in a similar way. Tables \ref{table:means_scanD} and \ref{table:means_scanE} show for scans D and E, respectively, the median $\alpha_m B$ and $B$ for each scenario. In addition, the median transverse and longitudinal components, $B_\perp$ and $B_\parallel$, as well as $\alpha_m B_\perp$, and $\alpha_m B_\parallel$, are also shown. First of all, the median $B$, $B_\perp$, and $B_\parallel$ values under S2 are all significantly lower than the equivalent values reported by \cite{Campbell2021a}. These weak average values underline that the target region is IN in nature. Further, as these fields can be understood as firmly in the weak field regime, where the circular polarization amplitudes are proportional to $B$ (and linear polarization amplitudes are proportional to $B^2$), recording lower median values in these parameters is {sensible on the inclusion of weaker Stokes profiles} that are accessed in the approximate range $2-2.4\times10^{-3}$ $I_c$ that could not be confidently measured in 2019. The median $\alpha_m$ values are very small, typically at $0.04$. However, this is a consequence of the shape of the distribution - the mean value is an order of magnitude bigger typically at $0.4-0.5$. The ratio $B_\perp / B_\parallel$ has been reported in other studies \citep{suarez2012,Campbell2021a} and is in fact slightly larger than 2019, indicating that as we increase the S/N, and access weaker fields, the magnetic field is getting more horizontal on average. {Since the ratio $B_\perp / B_\parallel$ would be $\pi/2$ for an isotropic distribution, this also indicates that accessing weaker fields has returned a distribution that is further from the isotropic case.} There is also a clear difference between scans D and E, with stronger $B$, $B_\parallel$, and $B_\perp$ values returned by SIR in all scenarios in scan E relative to scan D. It is also clear from the ratios that scan E is slightly more longitudinal on average than scan D. Finally, there is a clear difference between S2 and S3 for both scans, with the median $B$ values being about $20$ G stronger when the weakest linear polarization parameters are not eliminated. Further, the $B_\perp / B_\parallel$ ratio is slightly larger in S3 compared to S2 in both scans, and the median $\alpha_m$ values are unchanged. 

\begin{table}
\centering
\caption{Unsigned median magnetic field strengths and flux densities, and its horizontal (transverse) and vertical (longitudinal) components, for scan D. The median $B$ is measured across all pixels with at least one polarization parameter with maximum amplitude $>5\sigma_n$ across the $15648.5$ $\mathrm{\AA}$ line, but when computing the corresponding $B_\perp$ and $B_\parallel$ component only profiles that had Stokes $Q$ or $U$ $> 5\sigma_n$ in the former case, and Stokes $V$ $> 5\sigma_n$ in the latter case, are included.}       
\label{table:means_scanD}      
\begin{tabular}{|c | c | c | c| c| c| c|}

\cline{2-4}
   \multicolumn{1}{c|}{} & \multicolumn{1}{c|}{Scenario 1} & \multicolumn{1}{c|}{Scenario 2} &
    \multicolumn{1}{c|}{Scenario 3}\\

\hline
  $B$ & 206.4 G & 119.3 G & 140.6 G \\
 
  $B_\parallel$ & 92.6 G & 71.4 G & 77.9 G \\

  $B_\perp$ & 200.9 G & 153.2 G & 188.4 G \\

  $\alpha_m$ & 0.04 & 0.04 & 0.04 \\
  
  $\alpha_m B$ & 7.3 G & 4.7 G & 5.6 G \\
  
  $\alpha_m B_\parallel$ & 3.3 G & 2.8 G & 3.1 G \\
  
  $\alpha_m B_\perp$ & 7.1 G & 6.0 G & 7.5 G \\
  
  $B_\perp / B_\parallel$ & 2.2 & 2.1 & 2.4 \\

\hline
\end{tabular}
\end{table}

\begin{table}
\centering
\caption{As in Table~\ref{table:means_scanD}, but for scan E.}       
\label{table:means_scanE}      
\begin{tabular}{|c | c | c | c| c| c| c|}

\cline{2-4}
   \multicolumn{1}{c|}{} & \multicolumn{1}{c|}{Scenario 1} & \multicolumn{1}{c|}{Scenario 2} &
    \multicolumn{1}{c|}{Scenario 3}\\

\hline
  $B$ & 233.3 G & 163.6 G & 182.8 G\\
 
  $B_\parallel$ & 107.5 G & 98.2 G & 101.8 G\\

  $B_\perp$ & 221.1 G & 182.3 G & 220.1 G\\

  $\alpha_m$ & 0.04 & 0.04 & 0.04\\
  
  $\alpha_m B$ & 9.7 G & 6.0 G & 8.2 G \\
  
  $\alpha_m B_\parallel$ & 4.5 G & 3.6 G & 4.6 G\\
  
  $\alpha_m B_\perp$ & 9.2 G & 6.7 G & 9.9 G\\
  
  $B_\perp / B_\parallel$ & 2.1 & 1.9 & 2.2 \\
  
\hline
\end{tabular}
\end{table}

\begin{table}
    \caption{Percentage of pixels with inclinations in given ranges as determined by SIR under the three scenarios for scan D. The percentages are computed relative to the total number of pixels with at least one Stokes profile with maximum amplitude $>5\sigma_n$ across the $15648.5$ $\mathrm{\AA}$ line.}       
    \label{table:inclin_scanD}      
    \centering                          
    \begin{tabular}{|c |c | c | c| c|} 
    \cline{3-5}
    \multicolumn{1}{c}{} & & \multicolumn{3}{c|}{Scenario} \\
    \hline
      classification & range [$^\circ$] & 1 $[\%]$ & 2 $[\%]$ & 3 $[\%]$\\   
    \hline 
      highly vertical & $\gamma < 16$ & 3.9 & 16.6 & 16.0\\
     
      highly vertical & $\gamma > 164$ & 4.5 & 17.5 & 17.2\\
    
      intermediate & $15 < \gamma < 75$ & 33.5 & 25.1 & 24.6\\
      
      intermediate & $105 < \gamma < 165$ & 35.7 & 25.2 & 25.5\\
      
      highly inclined & $74 < \gamma < 106$ & 22.4 & 15.6 & 16.7\\
      
    \hline
    \end{tabular}
\end{table}

\begin{table}
    \caption{As in Table \ref{table:inclin_scanE}, but for scan E.}       
    \label{table:inclin_scanE}      
    \centering                          
    \begin{tabular}{|c |c | c | c| c|} 
    \cline{3-5}
    \multicolumn{1}{c}{} & & \multicolumn{3}{c|}{Scenario} \\
    \hline
      classification & range [$^\circ$] & 1 $[\%]$ & 2 $[\%]$ & 3 $[\%]$\\   
    \hline 
      highly vertical&$\gamma < 16$ & 5.0 & 19.0 & 18.3  \\
     
      highly vertical&$\gamma > 164$& 5.3 & 19.5 & 19.2 \\
    
      intermediate&$15 < \gamma < 75$& 33.9 & 23.1 & 23.0 \\
      
      intermediate&$105 < \gamma < 165$& 35.7 & 24.2 & 24.0 \\
      
      highly inclined&$74 < \gamma < 106$ & 20.2 & 14.1 & 15.5 \\
      
    \hline
    \end{tabular}
\end{table}

\subsection{Inclinations {and azimuths}}
The $\gamma$ distributions in Fig.~\ref{fig:inv_parameters} differ significantly between S1 on one hand and S2 and S3 on the other. The shape of the $\gamma$ distribution in S1 resembles the distributions derived from semi-empirical models produced by \cite{Borrero2011} with large noise thresholds ($>4.5\sigma_n$) specifically in terms of the small dip at $90^\circ$. As the authors explain, this dip is a consequence of asserting a large noise threshold in the weak field regime. Thus, by setting the threshold at $5\sigma_n$, the dip in the $\gamma$ distribution is created by excluding the weakest fields. This dip is also observed in the distribution from MURaM simulations by \cite{Campbell2021b}. For S2 and S3, the distribution is similar to \cite{Campbell2021a} in that large peaks are recorded at $0$, $90$, and $180^\circ$, which is a consequence of setting noisy Stokes $Q$, $U$, and $V$ profiles to zero. For instance, a Stokes vector with only Stokes $Q$ above the noise threshold and Stokes $V$ set to zero would be expected to return a $\gamma$ value close to $90^\circ$ by SIR when a good fit is achieved, {which is the reason for the central peak in the $\gamma$ distribution.} However, if this profile had a noisy Stokes $V$ included when inverted, the $\gamma$ would be more likely to show a more intermediate inclination.

Tables \ref{table:inclin_scanD} and \ref{table:inclin_scanE} classify the inclinations in terms of highly vertical, highly inclined, and intermediately inclined fields. The difference compared to 2019 is in the intermediately inclined fields, which have a much higher population. \cite{Campbell2021a} report over $70\%$ of pixels had a highly vertical classification, with a minority of pixels having a significant transverse component (i.e. classified as either intermediate or highly inclined). In scans D and E, remarkably, the situation is reversed - only $34.1\%$ in scan D and $38.5\%$ in scan E under S2 have a highly vertical classification, with the majority having a significant transverse component. The explanation for this has already been discussed in Sect.~\ref{sect:polarization}: there are a much larger number of pixels with both LP and CP in scans D and E. The very close similarity between S2 and S3 indicates that eliminating the weaker linear polarization parameter does not result in an altered distribution of $\gamma$ values, i.e. only the stronger linear polarization parameter is necessary to retrieve $\gamma$ (in addition to Stokes $V$, if a value that differs from close to $90^\circ$ is to be returned). Of course, there is a difference in $\phi$. The reality is that regardless of the scenario, since both Stokes $Q$ and $U$ must be measured confidently to constrain $\phi$, and so few pixels have both, retrieving $\phi$ (even without disambiguation) in a statistical sense is not feasible in this data. {Nevertheless, the peaks in the $|\cos\phi|$ distribution at $0.7$ and $1$ is a consequence of setting one of Stokes $Q$ or $U$ to zero while the other exceeds the $5\sigma$ threshold, forcing $\phi$ to be such that $\cos\phi$ is returned as $-1$, $-0.7$, $0.7$, or $1$ for $\phi$ values of $0$, $45$, $135$, or $180^\circ$ (or indeed these values shifted by $180^\circ$ due to a lack of disambiguation).} Finally, the clear difference between scans D and E is elucidated again, with slightly larger numbers of highly vertical classifications in scan E relative to scan D, and indeed slightly larger numbers of intermediate and highly inclined classifications in scan D relative to scan E.

\subsection{Case studies}\label{sect:case_studies}
\begin{figure*}
    \centering
    \includegraphics[width=.9\textwidth]{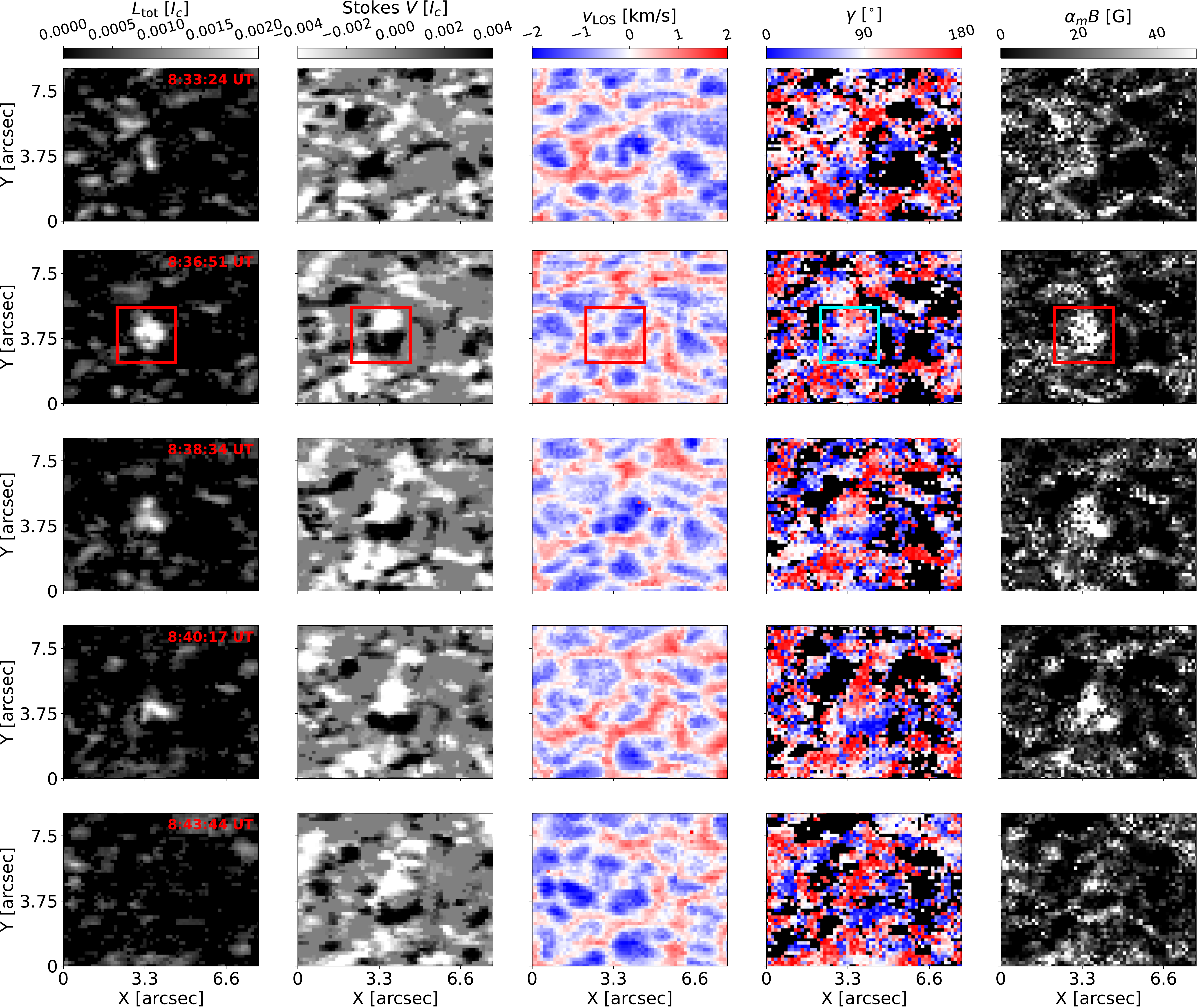}
    \caption{Case study of a magnetic loop in scan D. Shown from left to right is the $L_{\mathrm{tot}}$, Stokes $V$ at $15648.36$ $\mathrm{\AA}$ in the blue lobes of the $g_{\mathrm{eff}} = 3$ line, $v_{\mathrm{LOS}}$, $\gamma$, and $\alpha_m B$. The last three parameters are derived from SIR inversions under S2. The rows show subsequent frames with the time-stamp in the upper-right corner of the left-most plot. The box (\textit{solid line}) highlights the spatio-temporal location {of the distinct magnetic loop}.}
    \label{fig:ROI1}
\end{figure*}
\begin{figure}
    \centering
    \includegraphics[width=.8\columnwidth]{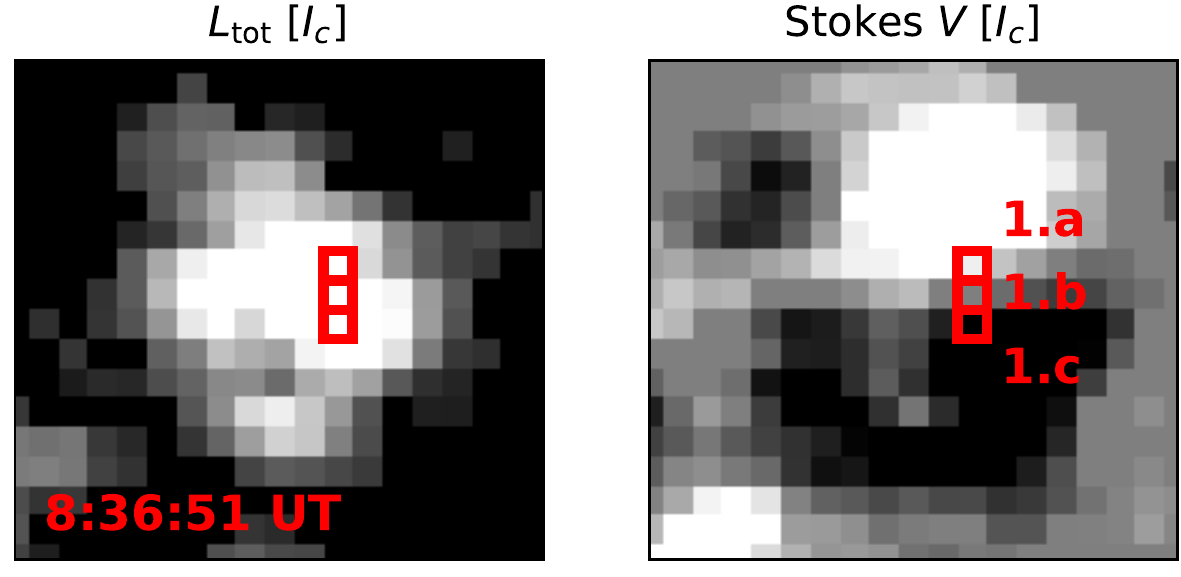}
    \caption{Close-up of the polarization in the boxed region in Fig.~\ref{fig:ROI1}, showing a distinct magnetic loop. The total linear polarization is shown in the \textit{left panel} and Stokes $V$ is shown in the \textit{right panel}. The full Stokes vectors of the three outlined pixels are shown in Fig.~\ref{fig:ROI1_profiles}. {The pixels are separated by $0.188''$ each.}}
    \label{fig:ROI1_loop}
\end{figure}

\begin{figure*}
    \centering
    \includegraphics[width=.9\textwidth]{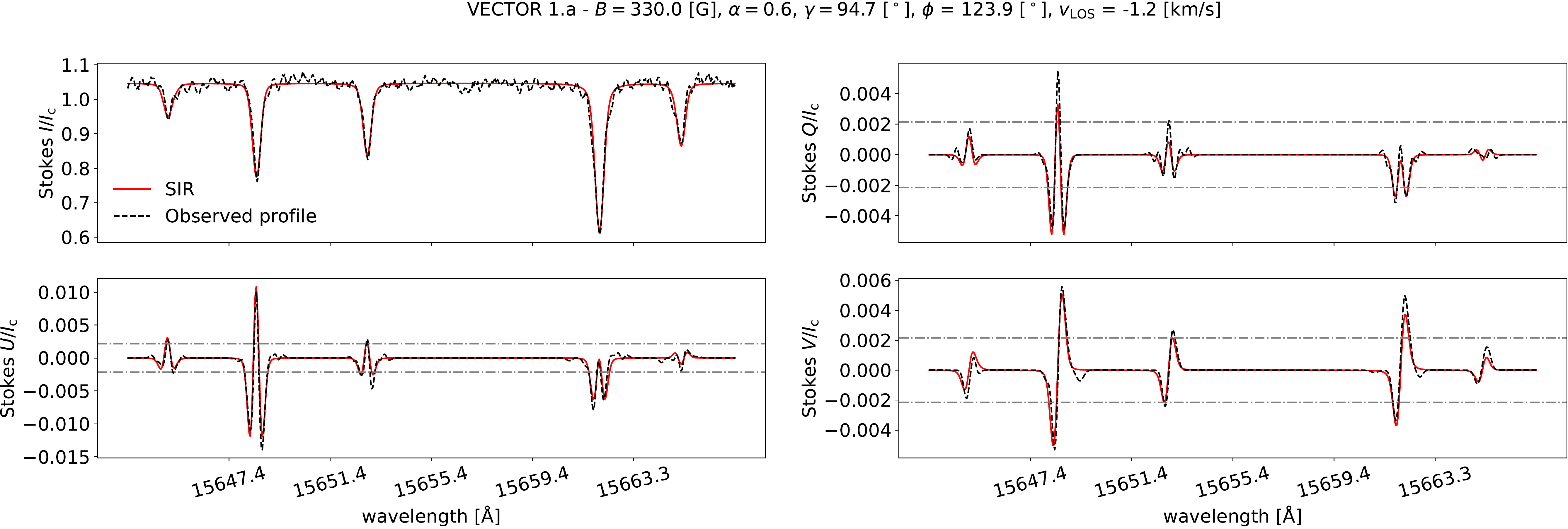}\vspace{.1cm}
    \includegraphics[width=.9\textwidth]{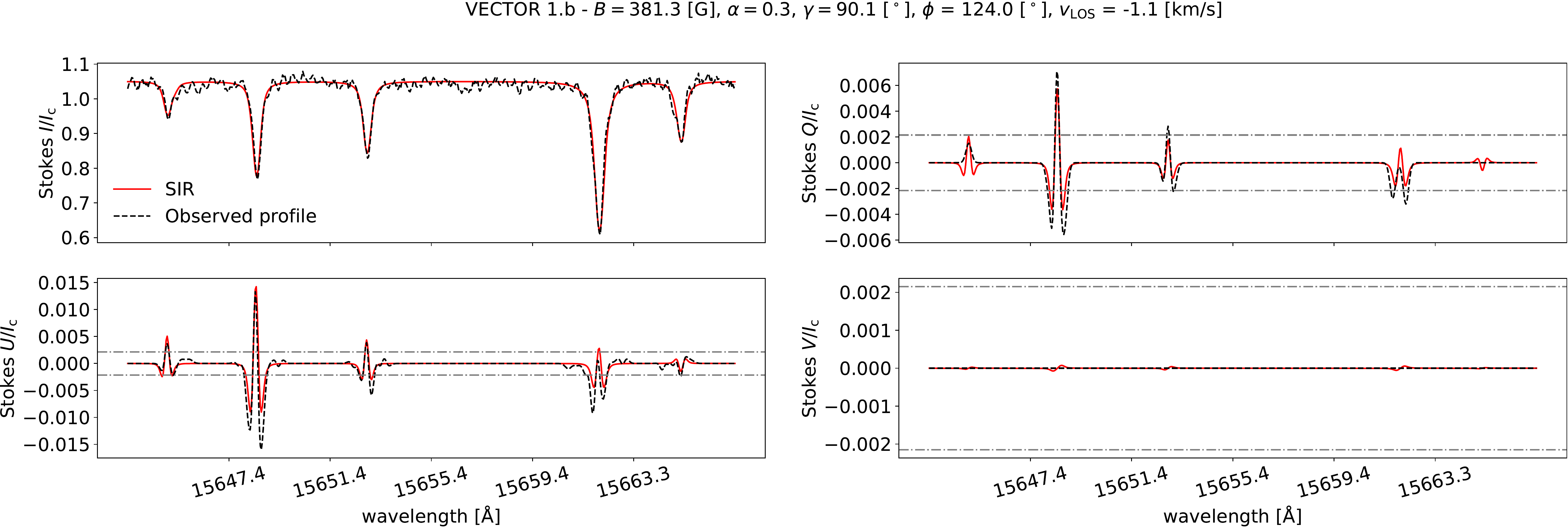}\vspace{.1cm}
    \includegraphics[width=.9\textwidth]{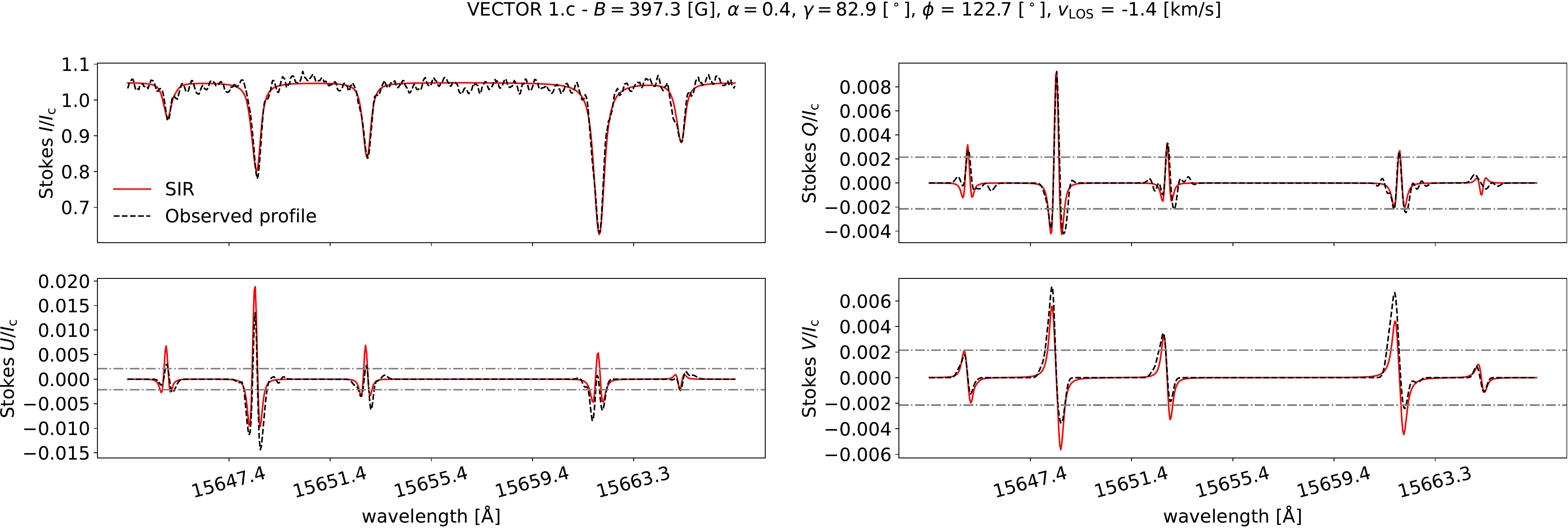}
    \caption{Full observed Stokes vectors 1.a, 1.b, and 1.c (\textit{black, dashed lines}), along with the corresponding synthetic vectors derived from SIR inversions (\textit{red, solid lines}). Stokes $Q$, $U$, and $V$ have been PCA-RVM reconstructed. The horizontal (\textit{dot-dashed}) lines show the $5\sigma_n$ noise thresholds. The locations of vectors 1.a, 1.b, and 1.c are shown in Fig.~\ref{fig:ROI1_loop}.  }
    \label{fig:ROI1_profiles}
\end{figure*}

A key strength of the GRIS-IFU lies in its ability to image the dynamics of small-scale magnetic features with high S/N, spatial resolution, and spectral resolution simultaneously. We now present two case studies and in order to make the analysis more efficient, we have developed an open-source application, SIR Explorer (SIRE), that enables users to navigate the multi-dimensional input and output files of SIR inversions. This tool is described in Appendix~\ref{sect:SIRE} and is used throughout this section. Figure~\ref{fig:ROI1} shows a magnetic loop visible in scan D. Figure~\ref{fig:ROI1_loop} shows a close-up of the polarization in the boxed region in Fig.~\ref{fig:ROI1}. Here we define the total linear polarization as,
\begin{equation}
    {L_{\mathrm{tot}}} = \frac{\int_{\lambda_b}^{\lambda_r} [Q^2(\lambda)+U^2(\lambda)]^{\frac{1}{2}}  d\lambda}{I_c \int_{\lambda_b}^{\lambda_r} d\lambda}.
\end{equation}
Figure~\ref{fig:ROI1} shows a magnetic feature whose structure resembles a magnetic loop. There is a large patch of linear polarization, which bridges two patches (or foot-points) of opposite polarity vertical fields. {As seen from the Dopplergram, }the linear polarization appears at the centre of the granule while the circular polarization is located in the intergranular lanes (IGLs). Over time, the circular polarization remains cemented in the lanes as the granule evolves. As the linear polarization vanishes below our detection capabilities, the two patches of circular polarization no longer appear to be in close contact. This magnetic feature has a lifetime of less than $10$ minutes. This shows how the evolution of the magnetic loop is influenced by the granular evolution. There are many pixels in this structure which have all three polarization profiles above the $5\sigma_n$ threshold. Figure~\ref{fig:ROI1_profiles} shows three example Stokes vectors, whose spatio-temporal locations are outlined in Fig.~\ref{fig:ROI1_loop}, across the PIL of the magnetic loop. The upper-most vector, vector 1.a, has a positive polarity but the field is highly inclined ($\gamma = 95^{\circ}$) and stronger than the median ($B = 330$ G, $\alpha_m B = 198$ G). The middle vector, vector 1.b, has no confidently measured Stokes $V$ profile. The pixel is located in the PIL and thus {the physical reason for the lack of a circular polarization profile could be either due to mixing of opposite polarities within the spatio-temporal resolution element or the magnetic vector could be purely transversal.} The field is still strong ($B = 381$ G), but the filling factor is halved compared to profile 1.a ($\alpha_m = 0.3$) and thus the $\alpha_m B$ is significantly lower ($\alpha_m B = 114$ G). Due to the lack of a confidently measured Stokes $V$ signal, we cannot access the longitudinal component of the field, and thus the $\alpha_m B$ is likely under-estimated in vector 1.b. Finally, in the pixel below the PIL, vector 1.c, the polarity of the field is negative but the field is still strong ($B = 397$ G, $\alpha_m B = 159$ G) and highly inclined ($\gamma = 83^{\circ}$). All three Stokes vectors have very strong Stokes $Q$ and $U$ signals, and thus there is information about $\phi$ available in each profile. The $\phi$ values are very consistent, differing by only $1-2^{\circ}$. Of course, the ambiguity in the $\phi$ remains, so although $\phi$ values of $123-124^{\circ}$ are returned by SIR, values of $303-304^{\circ}$ would also be equally acceptable solutions. Of course, as all three pixels are located in the granule their $v_{\mathrm{LOS}}$ values are consistently negative.

\begin{figure*}
    \centering
    \includegraphics[width=.9\textwidth]{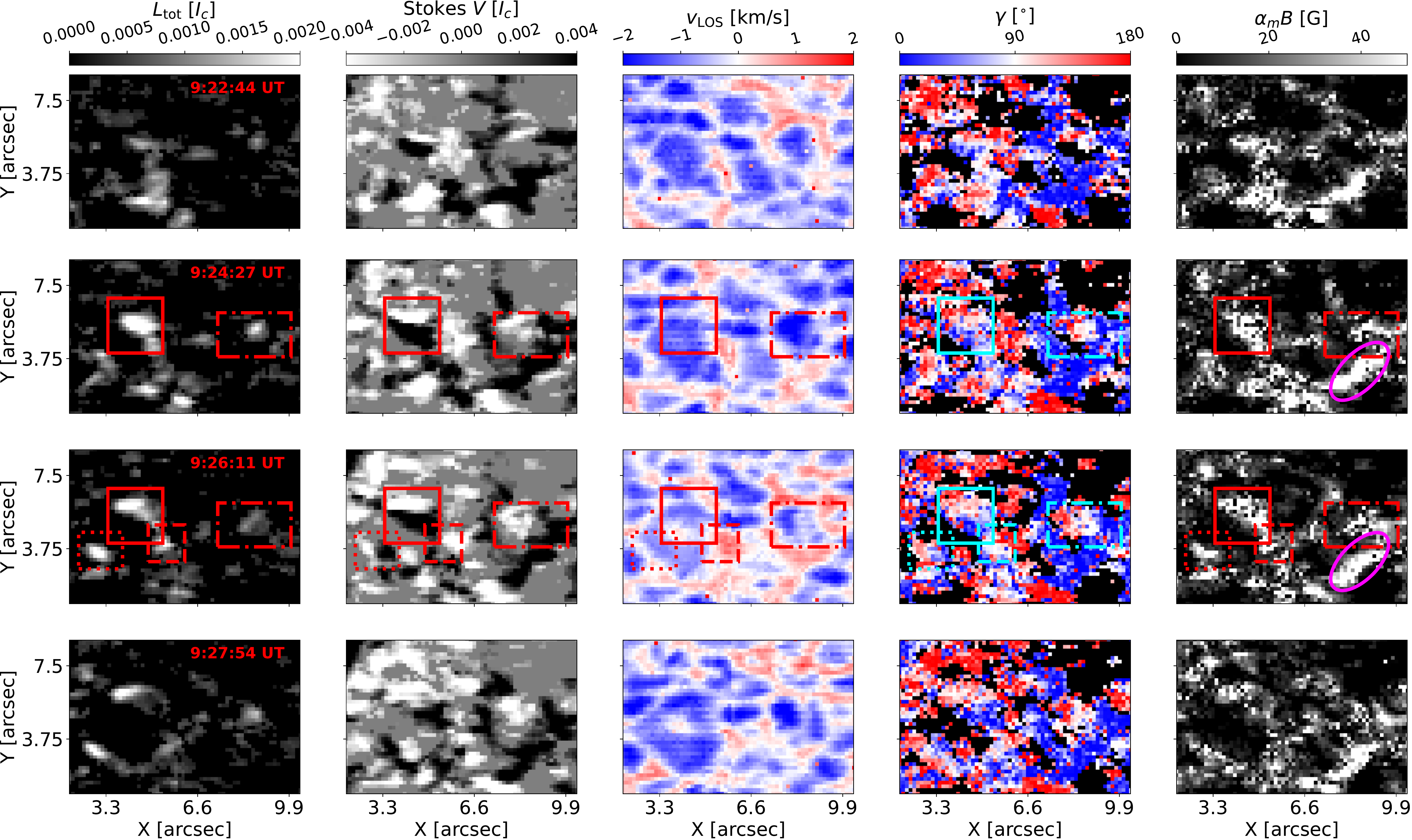}
    \caption{Case study of a series of magnetic loops in scan E. The plots follow the same layout as in Fig.~\ref{fig:ROI1}. The \textit{solid}, \textit{dashed}, \textit{dotted} and \textit{dot-dashed} \textit{red} boxes highlight the locations of magnetic loops. The \textit{solid, magenta} ellipse highlights the location of a strong longitudinal patch of magnetic flux. }
    \label{fig:ROI2}
\end{figure*}

\begin{figure}
    \centering
    \includegraphics[width=.8\columnwidth]{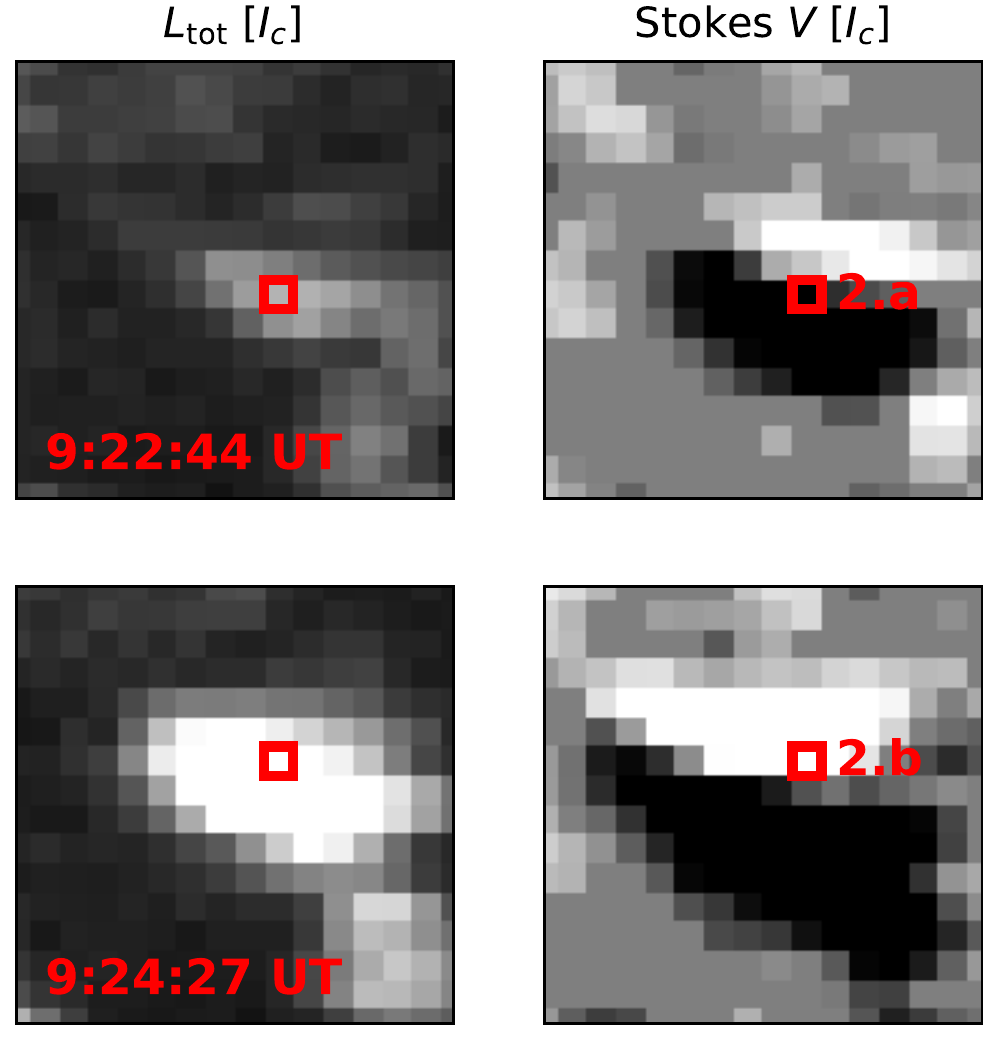}
    \caption{As in Fig~\ref{fig:ROI1_loop}, but for a magnetic loop outlined by the solid box in Fig.~\ref{fig:ROI2}. The full Stokes vectors of the two outlined pixels are shown in Fig.~\ref{fig:ROI2_profiles}.}
    \label{fig:ROI2_loop}
\end{figure}

\begin{figure*}
    \centering
    \includegraphics[width=.9\textwidth]{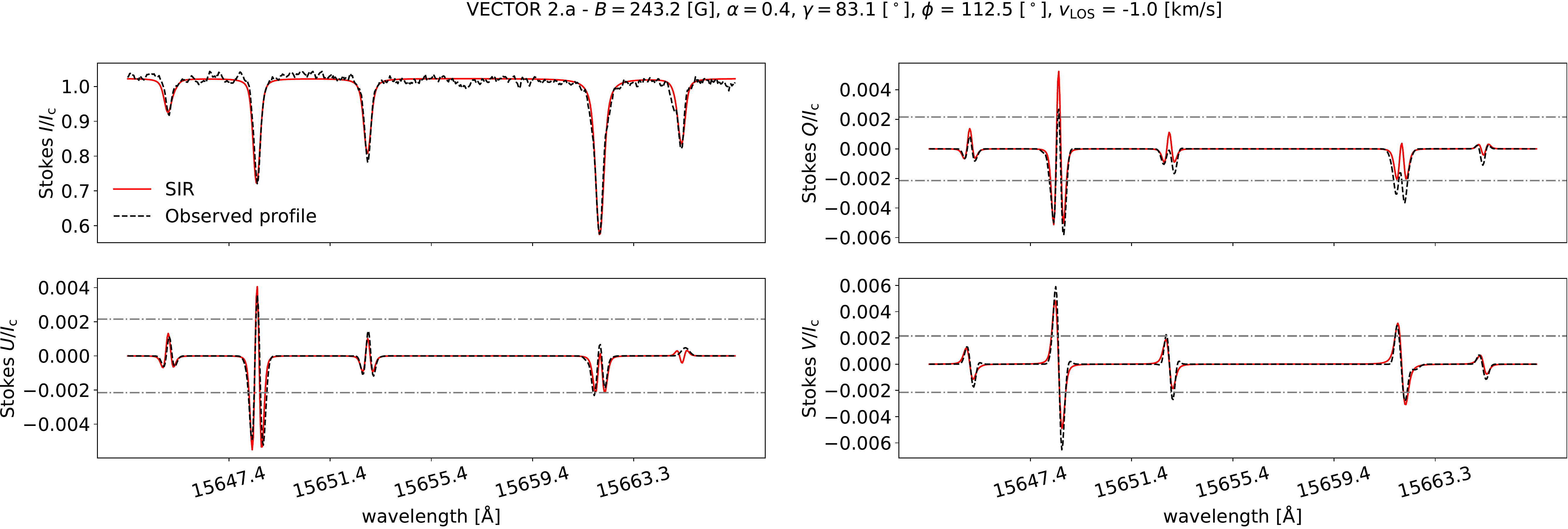}\vspace{.5cm}
    \includegraphics[width=.9\textwidth]{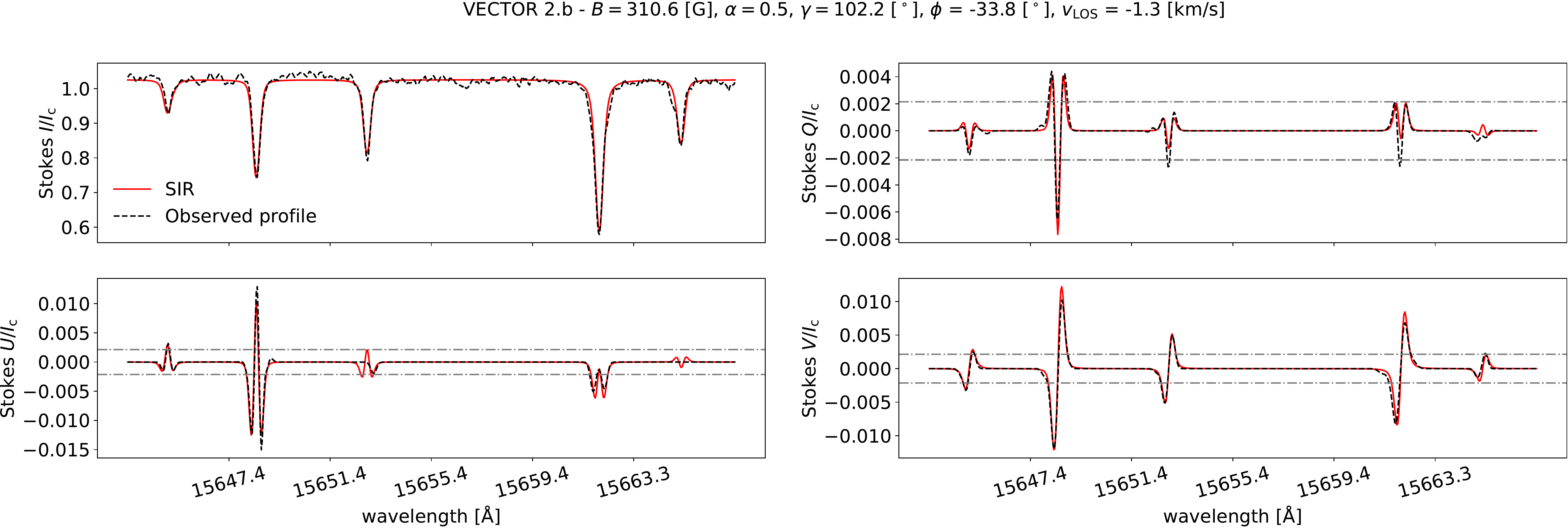}
    \caption{As in Fig.~\ref{fig:ROI1_profiles}, but for vectors 2.a and 2.b. The spatio-temporal locations of the pixels are shown in Fig.~\ref{fig:ROI2_loop}. }
    \label{fig:ROI2_profiles}
\end{figure*}

Figure~\ref{fig:ROI2} shows the second case study of {a series of magnetic loops }present in scan E. In this region, several magnetic loops, whose locations are highlighted with boxes in Fig.~\ref{fig:ROI2}, are present and appear to form a continuous ``serpentine'' structure. {Alternatively, these could be independent loops formed by the action of a small-scale dynamo. }Most of the magnetic flux is located in IGLs, with linear polarization present in the PIL which is most commonly located at the granule-IGL boundary or in the granule. A close-up of the magnetic loop highlighted by the solid box in Fig.~\ref{fig:ROI2} is shown in Fig.~\ref{fig:ROI2_loop} for two frames. Since opposite polarity Stokes $V$ profiles cancel in the PIL, the most common location to find pixels with all three polarization parameters above the noise threshold is one pixel adjacent to the PIL. For instance, vectors 2.a and 2.b, shown in Fig.~\ref{fig:ROI2_profiles} with their spatio-temporal location shown in Fig~\ref{fig:ROI2_loop}, are located on either side of the PIL and thus have opposing polarities ($\gamma = 83^\circ$ versus $\gamma = 102^\circ$). The $B$ and $\alpha_m B$ values of profile 2.a are smaller ($B = 243$ G, $\alpha_m B = 97$ G) than profile 2.b ($B = 311$ G, $\alpha_m B = 155$ G) as the magnetic loop increases in strength as it emerges in the frame from which profile 2.a is selected and reaches its peak in the frame from which profile 2.b is found. Both of these pixels are located in granular upflows. 

\begin{figure}
    \centering
    \vspace{0.3cm}
    \includegraphics[width=.9\columnwidth]{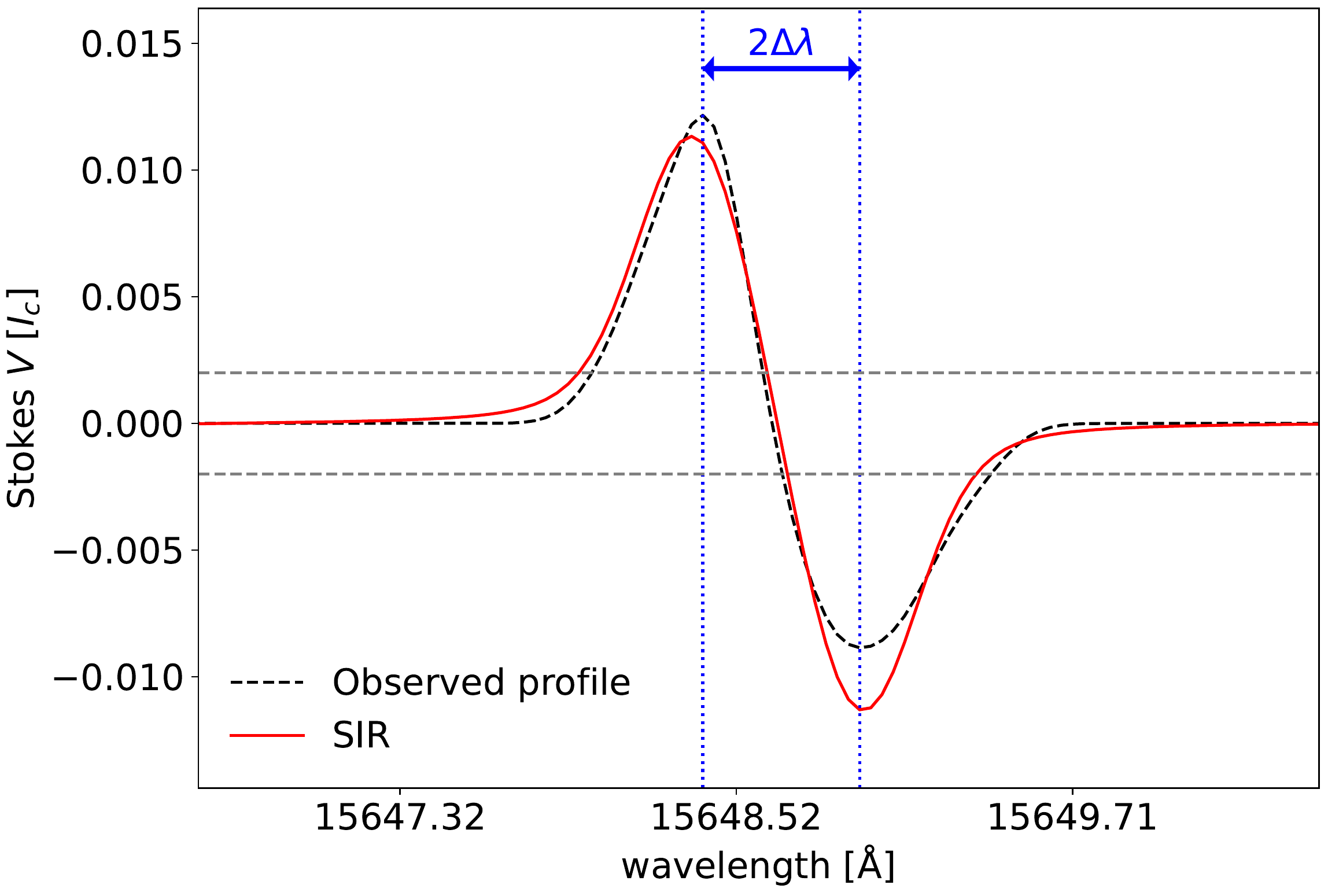}
    \caption{Sample Stokes $V$ profile for a pixel located in the strong magnetic element in case study 2. The \textit{dotted, blue} lines show the wavelength positions of the blue and red lobes for the most magnetically sensitive line, from which the strong field approximation is used to estimate $B_\parallel = 812$ G. The synthetic profile from SIR is also shown, which has model parameters of $B = 854$ G, $\alpha_m = 0.14$, $\gamma = 0^\circ$, $v_{\mathrm{LOS}} = 2.2$ km/s.}
    \label{fig:SFA}
\end{figure}

Considering the wider context of the second case study, the series of magnetic loops are connected to a much stronger longitudinal magnetic element shown in the bottom-right of Fig.~\ref{fig:ROI2} in the ellipse.
The temporal evolution of this patch of longitudinal flux can be traced throughout the full $1$ hour time series and therefore persists even as the much shorter-lived magnetic loops appear and disappear. This magnetic element is first visible at the start of the time series further to the right, and is migrated and coalesced by granular evolution until it covers a smaller surface area by the end of the scan. The presence of this strong longitudinal flux element is reminiscent of the magnetic loop reported by \cite{Campbell2021a}, which was also located next to a strong longitudinal magnetic element. For normal Zeeman triplets, with an effective Land\'e g-factor, $g_{\mathrm{eff}}$, characterized by two split $\sigma$ components and an unshifted $\pi$ component, the two $\sigma$ components are separated from the rest wavelength, $\lambda_0$, by
\begin{equation}\label{eqn:SFA}
    \Delta\lambda = \pm \frac{e}{4\pi m_e c}\lambda_0^2 g_{\mathrm{eff}}B \approx \pm4.67\times10^{-13} \lambda_0^2 g_{\mathrm{eff}}B,
\end{equation}
where $\Delta \lambda$ and $\lambda_0$ are in units of $\textrm{\AA}$ and $B$ is in units of G, $c$ is the speed of light, $m_e$ is the mass of an electron, and $e$ is the electron charge. In what is known as the strong field approximation (SFA), by measuring $\Delta \lambda$ it is therefore possible to measure $B$ directly from the splitting of the lobes of Stokes $V$ when the field is strong enough \citep{khomenko2003, Nelson2021}. SIRE provides a simple calculator which allows users to quickly calculate an estimate of $B$ based on the separation of the lobes, which can also be measured using the wavelength slider. Figure \ref{fig:SFA} shows a sample Stokes $V$ profile from the structure. The Stokes $V$ profile is asymmetric, particularly in the red $\sigma$ lobe, but SIR can still produce an estimate of $B$ even without introducing the gradients in $v_{\mathrm{LOS}}$ or $B$ that would be necessary to better fit this profile. The pixel had no linear polarization. The values returned from SIR were: $B = 854$ G, $\alpha_m = 0.14$, $\gamma = 0^\circ$, $v_{\mathrm{LOS}} = 2.2$ km/s. The SFA estimate is returned as $B_{\mathrm{SFA}} = 812$ G. The wavelength positions from which this estimate was measured is indicated in Fig.~\ref{fig:SFA}. SIR must determine $B$ such that a good fit is found for all lines, and as such it is not abnormal that there is a small difference in the measurements, especially given that the wavelength position of the synthetic and observed blue $\sigma$ lobe differ very slightly. In any case, a shift of a single increment in wavelength is equivalent to $58$ G, and so the values are within an acceptable uncertainty interval of each other, and serves as a sanity check demonstrating that the inversions are well calibrated.

\section{Discussion}\label{sect:discussion}

The quest to reveal higher fractions of the FOV as displaying confidently measured Zeeman-induced polarization signal continues. The fact that $65-67\%$ of the IN FOV has confidently measured polarization {in at least one polarization parameter} is a remarkable feat for the GREGOR/GRIS-IFU, {however linear and circular polarization is only confidently detected simultaneously in $16-18\%$.} The increase in $\delta I_{\mathrm{rms}}$ could be the result of having achieved closer to diffraction limited resolution. Indeed, the exceptional seeing conditions during scans D and E could be responsible for this. However, it is certainly the case that these observations have not achieved the diffraction limited spatial resolution that GREGOR is capable of, because the spatial sampling of the GRIS-IFU in the $y-$direction, even in double sampling mode, is insufficient. The increase in $\delta I_{\mathrm{rms}}$ could also be the result of a small reduction in stray light. Another possibility for the large polarization fractions recorded is that we were lucky with the target, and these explanations are not mutually exclusive.

{\cite{lites2008} report a mean vertical apparent flux density of $11$ G. This value, determined from Hinode/SP observations of the Fe I $6301/2$ $\mathrm{\AA}$ line pair, was found assuming $\alpha_m = 1$. However, the authors themselves recommend that this value should be reduced by $30\%$ to $7.7$ G, before comparison with other works, as the assumption that $\alpha_m = 1$ is almost certainly untrue. The very large fraction of the FOV in scans D and E which have a CP signal makes a comparison worthwhile. The corresponding average value from this work is the median $\alpha_m B_\perp$, listed in Table~\ref{table:means_scanD} and ~\ref{table:means_scanE} as between $2.8$ G and $4.6$ G. As their mean is calculated over the full FOV of the Hinode/SP scan, and their FOV is much larger and includes network patches, it is sensible that the value for the GREGOR/GRIS-IFU is smaller. Further as the NIR Fe I line pair is more magnetically sensitive than the visible line pair, it is expected that we record a smaller average value in our maps. However, the value is a function of the effective spatial resolution and S/N - indeed, in deep mode observations \cite{lites2008} return a distribution for the vertical apparent longitudinal flux density that peaks at a much lower value of $1.2$ G. Our value therefore rests comfortably between these two values. Aside from a comparison, it must be emphasised that all of these values could be under-estimates if there remains a substantial amount of unresolved mixed-polarities.}

The much larger number of pixels which have a LP signal in scans D and E, relative to 2019, allowed us to infer the inclination angle in a larger fraction of the FOV. Using SIR, we were then able to determine that a large majority ($>60\%$) of the pixels which had at least one Stokes parameter with a maximum signal greater than $5\sigma_n$ across the $15648.5$ $\mathrm{\AA}$ line were classifiable as either highly inclined or intermediately inclined. In other words, a majority of magnetised pixels displayed a significant transverse magnetic component of the magnetic field. This is because the number of intermediately inclined fields outnumbers both the highly inclined and highly vertical populations. The ratio $B_\perp / B_\parallel$ is larger than 2019, which indicates that as the S/N increased, and thus we were able to access weaker fields, the magnetic field is revealed as more transverse on average. Determining this ratio from observations is important because it is useful for optimisation of the initial and boundary conditions of radiative magnetohydrodynamic simulations \citep{steiner2008}. The question which naturally follows is whether this is an uncommon quirk of the specific observational target in scans D and E, or whether this is a glimpse of what we can expect to see when larger aperture telescopes like the Daniel K. Inouye Solar Telescope (DKIST: \cite{DKIST}) and European Solar Telescope (EST: \cite{EST}) take similar observations in the NIR lines with significantly higher effective spatial resolution. Importantly, this result was obtained even with a very conservative approach to noise tolerance that favours longitudinal inclinations. In scenario 2 (and 3), we set Stokes profiles which did not satisfy the $5\sigma_n$ threshold to zero, which means in many pixels the $\gamma$ is likely to be either $0^\circ, 90^\circ$, or $180^\circ$, and this is reflected in the large peaks at these values in the $\gamma$ distributions in Fig.~\ref{fig:inv_parameters}. {Of course, Stokes $Q$ and $U$ are set to zero in a much larger number of pixels than Stokes $V$.} {While our approach favours longitudinal inclinations, that does not mean our determination of $B_\perp / B_\parallel$ is biased in favour of $B_\parallel$. It must be stressed that the opposite is true - our determination of $B_\perp / B_\parallel$ is biased in favour of horizontal fields because $B_\perp$ is determined over a much smaller population of pixels with significantly stronger fields \citep{steiner2012}.} To determine an unbiased value for $B_\perp / B_\parallel$ it may be necessary to approach this problem by applying a threshold to each Stokes parameter separately whose amplitude is determined by the magnetic field strength, rather than applying the same threshold to Stokes $Q$, $U$, and $V$ as determined purely noise (i.e. by the amplitude of the $5\sigma_n$ threshold).

The magnetic loops shown in both case studies are clear demonstrations of how the photospheric magnetic field is organized in the QS IN. Of particular significance is the apparently serpentine nature of this magnetism. In terms of magnetic loops in the IN, there are a large number of statistical studies that report on the phenomenon, especially from the Imaging Magnetograph (IMaX)/Sunrise experiment \citep{danilovic_imax_2010,marian2012} and the SST \citep{gosic2021,gosic2022,ledvina2022}. Most of these studies focus on analysing bi-polar patches of circular polarization where cancellation could take place and where horizontal fields are expected to be found along the PIL. For instance, the SST observations by \cite{ledvina2022} with the Fe I $5576$ $\mathrm{\AA}$ and $6301$ $\mathrm{\AA}$ lines, and with a FOV $30$ times greater by area than the GRIS-IFU scans in this paper, found $38$ magnetic loops in a $42.5$ minute time series, but find no evidence of horizontal fields at the PIL in any of them. In terms of serpentine magnetism, \cite{harra2010} reported on observations of a serpentine magnetic field between two large bi-polar patches of magnetic flux in an emerging flux region. {However, their target was not an IN region.} We have revealed a serpentine structure in the QS IN for the first time with unambiguous linear polarization along the PIL. However, there is still a clear need for higher resolution observations. If one examines the circular polarization at the PIL, the mixing of opposite polarity signals within the spatio-temporal resolution element means full-vector spectropolarimetry remains elusive (see vector 1.b in Fig.~\ref{fig:ROI1_profiles}). Nevertheless, full-vector spectropolarimetry is achieved just one pixel adjacent in both case scenarios examined. Further there is a need for multi-wavelength facilities with spectral diagnostics sensitive to the upper photosphere and lower chromosphere, such as will be available at the DKIST \citep{dkistcsp}, the EST \citep{EST}, and also the Sunrise III balloon experiment when launched, which will allow us to assess whether these small-scale cancellation sites, which could be pervasive across the quiet solar surface, are capable of contributing to the heating of the lower chromosphere.


\appendix
\section{SIR Explorer}\label{sect:SIRE}
As spectropolarimetric solar datasets become ever larger and more complex, the multi-dimensional data cubes produced from observatories become increasingly non-trivial to analyse. Inversions of this multi-dimensional data increase in complexity too as the number of spectral lines that are inverted continues to increase irrespective of the data volume. To meet the challenge of inverting such large data cubes, {the parallelized Python wrapper \citep{ricardo2021} to both the SIR \citep{SIR} and the Departure coefficient aided Stokes Inversion based on Response functions (DeSIRe; \cite{ desire}) codes} enables users to easily spread the computational problem across many CPU nodes on high performance computing facilities. However, even when the data is inverted, the data products still need to be analysed and browsing the data products of inversions, of which there are many, becomes increasingly cumbersome. SIR Explorer (SIRE), is a Python $3.9$ graphical user interface (GUI) application that aims to make the analysis and exploration of inversion inputs and outputs associated with SIR/DeSIRe a little faster and a little easier \citep{SIRE}. 

\begin{figure*}
    \centering
    \includegraphics[width=\textwidth]{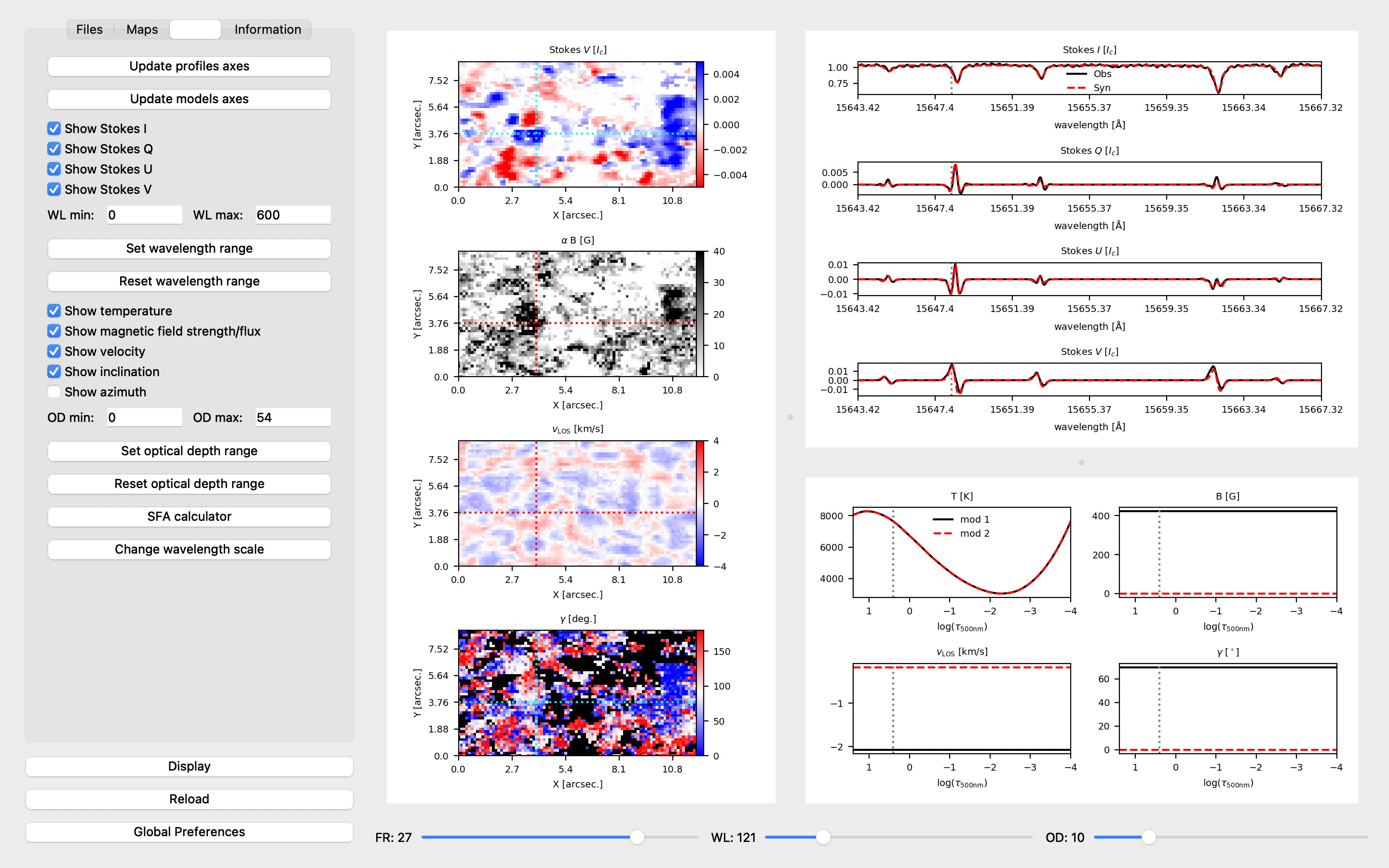}
    \caption{Main window of SIRE with scan D loaded into the application.}
    \label{fig:SIRE}
\end{figure*}

Figure \ref{fig:SIRE} shows the S2 inversions of scan D loaded into SIRE. The user interface (UI) of SIRE is split into three main regions:
\begin{enumerate}
    \item The control panel, located on the left of the main window, which allows the user to load datasets,
    \item The three canvases, upon which datasets are displayed, and
    \item The widget bar, underneath the canvases, which has three sliders to control the frame (FR), wavelength (WL), and optical depth (OD) indices.
\end{enumerate}
SIRE is designed to pick up where the parallelized Python wrapper to SIR/DeSIRe leaves off. SIRE's required input file structure, detailed in Table \ref{table:SIRE_input_files}, is therefore dictated by the output file structure of the wrapper. There are three mandatory files that are always required. The first mandatory file is the primary models output by SIR/DeSIRe, which are an array of the one-dimensional atmospheric files with a user-defined number of optical depth points, $n(\tau)$. The other two mandatory files are the observed profiles, which are the input Stokes vectors provided to SIR/DeSIRe, and the synthetic profiles, which are the final output Stokes vectors produced by SIR/DeSIRe when solving radiative transfer, which must both have the same number of wavelength points, $n(\lambda)$. The observed profiles do not need to be real observations. If inverting synthetic vectors produced from simulation snapshots, these may synthetic in origin. There are a number of optional files that can be provided. These include: secondary models, if SIR/DeSIRe is employed with two models, primary (and secondary) macroturbulence files which contains the macroturbulence of the model and also the $\alpha$ and stray light fraction values, and binary map(s), which are a user-defined array that may be provided to remove selected pixels from the maps of the magnetic parameters.

\begin{table}
\centering
\caption{SIRE's input files and the required array structure. The $\star$ symbol indicates that the file is mandatory in all circumstances. The $\diamond$ symbol indicates that the file is mandatory when two models are provided.}       
\label{table:SIRE_input_files}      
\begin{tabular}{|c | c | c |}

\hline
  Input file & Array shape ($t=1$) & Array shape ($t>1$)\\
\hline  
  Primary models $\star$ $\diamond$ & [11, $n(\tau)$, $y$, $x$] & [$t$, 11, $n(\tau)$, $y$, $x$] \\
  Observed profiles $\star$ $\diamond$ & [4, $n(\lambda)$, $y$, $x$] & [$t$, 4, $n(\lambda)$, $y$, $x$] \\
  Synthetic profiles $\star$ $\diamond$ & [4, $n(\lambda)$, $y$, $x$] & [$t$, 4, $n(\lambda)$, $y$, $x$]\\
  Secondary models $\diamond$ & [11, $n(\tau)$, $y$, $x$] & [$t$, 11, $n(\tau)$, $y$, $x$] \\
  Primary macroturbulence files  $\diamond$ & [3, $y$, $x$] & [$t$, 3, $y$, $x$] \\
  Secondary macroturbulence files & [3, $y$, $x$] & [$t$, 3, $y$, $x$]\\
  Binary map(s) & [$y$, $x$] & [$t$, $y$, $x$]  \\

\hline
\end{tabular}
\end{table}

There are three canvas objects: the maps where images of the Stokes profiles and model parameters may be displayed, the plots on the upper right of the interface of the observed and synthetic Stokes profiles for the selected pixel, and the plots on the bottom right of the interface of the model parameters as a function of optical depth for the selected pixel. Each of the canvases may be resized in proportion to one another, and each of them may be collapsed entirely. The control panel provides the user with the ability to determine exactly which parameters should be shown in each canvas. For instance, in Fig~\ref{fig:SIRE} only maps of Stokes $V$, $\alpha_m B$, $v_{\mathrm{LOS}}$, and $\gamma$ are selected and shown.

The most important functionality of SIRE concerns the way in which the user is able to navigate the dataset. The controls have been designed to be as simple and fast as possible. When a dataset is first loaded, the pixel with zeroth co-ordinates in each dimension ($t = 0$, $n(\lambda) = 0$, $y = 0$, $x = 0$, $n(\tau) = 0$) will be plot by default. The user can left-click any of the maps and the synthetic Stokes vector, observed Stokes vector, and model parameters for the corresponding pixel co-ordinates will be plotted. In addition, the user may adjust the selected pixel by using key-pressing the `UP', `DOWN', `LEFT', and `RIGHT' arrow keys on their keyboard. The location of the selected pixel is denoted by vertical and horizontal lines on the maps. The sliders in the widget bar allow the user to adjust the frame (FR), wavelength (WL), and optical depth (OD) indices at which the data is retrieved from the appropriate files. The user can left-click and drag the sliders. The code that updates the maps and plots will not be executed until the slider is released. The user can also key-press `Q' and `E' to decrease or increase, respectively, the frame index by $1$. Similarly, the user can key-press `A' and `D' for wavelength, or `Z' and `C' for optical depth.

\begin{acknowledgments}
{The authors would like to thank the anonymous referee whose feedback helped to significantly improve this manuscript.} We express our appreciation also to Carlos Dominguez-Tagle, whose work with the GRIS-IFU made these observations possible, and to all the engineering, operating, and technical staff at GREGOR for their assistance during the observing campaign, including Miguel Esteves Perez, Saida Milena Díaz Castillo, Karin Gerber, and Oliver Wiloth. Gratitude is extended to Juan Manuel Borrero and Lucia Kleint for their advice and insightful discussions. RJC thanks Robert Ryans for IT support and assistance in utilizing QUB's high performance computing (HPC) facilities. We thank Carsten Denker and Christoph Kuckein for assistance with operating the High-resolution Fast Imager (HiFI) instrument and associated data reduction. This research has received financial support from the European Union’s Horizon $2020$ research and innovation program under grant agreement No. $824135$ (SOLARNET). RJC acknowledges support from the Northern Ireland Department for the Economy (DfE) for the award of a PhD studentship. RJC and MM acknowledge support from the Science and Technology Facilities Council (STFC) under grant No. ST$/$P000304$/$1 $\&$ ST$/$T00021X$/$1. The $1.5-$meter GREGOR solar telescope was built by a German consortium under the leadership of the Leibniz-Institute for Solar Physics (KIS) in Freiburg with the Leibniz Institute for Astrophysics Potsdam, the Institute for Astrophysics Göttingen, and the Max Planck Institute for Solar System Research in Göttingen as partners, and with contributions by the Instituto de Astrofísica de Canarias and the Astronomical Institute of the Academy of Sciences of the Czech Republic. The redesign of the GREGOR AO and instrument distribution optics was carried out by KIS whose technical staff is gratefully acknowledged. Helioseismic and Magnetic Imager (HMI) magnetograms, courtesy of NASA/SDO and the AIA, EVE, and HMI science teams, were used during observations for target selection. This work was supported by Fundação para a Ciência e a Tecnologia (FCT) through the research grants UIDB/04434/2020 and UIDP/04434/2020.
\end{acknowledgments}

%

{\facilities{GREGOR solar telescope \citep{Schmidt2012, kleint2020}.}}


{\software{\hyperlink{https://github.com/r-j-campbell/SIRExplorer}{SIR explorer}, \citep{SIRE}, Astropy \citep{astropy3}, Matplotlib \citep{mpl}, Numpy \citep{numpy}.}}


\bibliography{sample631}{}

\begin{thebibliography}{}
\expandafter\ifx\csname natexlab\endcsname\relax\def\natexlab#1{#1}\fi
\providecommand{\url}[1]{\href{#1}{#1}}
\providecommand{\dodoi}[1]{doi:~\href{http://doi.org/#1}{\nolinkurl{#1}}}
\providecommand{\doeprint}[1]{\href{http://ascl.net/#1}{\nolinkurl{http://ascl.net/#1}}}
\providecommand{\doarXiv}[1]{\href{https://arxiv.org/abs/#1}{\nolinkurl{https://arxiv.org/abs/#1}}}

\bibitem[{{Asplund} {et~al.}(2009){Asplund}, {Grevesse}, {Sauval}, \&
  {Scott}}]{asplund}
{Asplund}, M., {Grevesse}, N., {Sauval}, A.~J., \& {Scott}, P. 2009, \araa, 47,
  481, \dodoi{10.1146/annurev.astro.46.060407.145222}

\bibitem[{{Astropy Collaboration} {et~al.}(2022){Astropy Collaboration},
  {Price-Whelan}, {Lim}, {Earl}, {Starkman}, {Bradley}, {Shupe}, {Patil},
  {Corrales}, {Brasseur}, {N{\"o}the}, {Donath}, {Tollerud}, {Morris},
  {Ginsburg}, {Vaher}, {Weaver}, {Tocknell}, {Jamieson}, {van Kerkwijk},
  {Robitaille}, {Merry}, {Bachetti}, {G{\"u}nther}, {Aldcroft},
  {Alvarado-Montes}, {Archibald}, {B{\'o}di}, {Bapat}, {Barentsen},
  {Baz{\'a}n}, {Biswas}, {Boquien}, {Burke}, {Cara}, {Cara}, {Conroy},
  {Conseil}, {Craig}, {Cross}, {Cruz}, {D'Eugenio}, {Dencheva}, {Devillepoix},
  {Dietrich}, {Eigenbrot}, {Erben}, {Ferreira}, {Foreman-Mackey}, {Fox},
  {Freij}, {Garg}, {Geda}, {Glattly}, {Gondhalekar}, {Gordon}, {Grant},
  {Greenfield}, {Groener}, {Guest}, {Gurovich}, {Handberg}, {Hart},
  {Hatfield-Dodds}, {Homeier}, {Hosseinzadeh}, {Jenness}, {Jones}, {Joseph},
  {Kalmbach}, {Karamehmetoglu}, {Ka{\l}uszy{\'n}ski}, {Kelley}, {Kern},
  {Kerzendorf}, {Koch}, {Kulumani}, {Lee}, {Ly}, {Ma}, {MacBride}, {Maljaars},
  {Muna}, {Murphy}, {Norman}, {O'Steen}, {Oman}, {Pacifici}, {Pascual},
  {Pascual-Granado}, {Patil}, {Perren}, {Pickering}, {Rastogi}, {Roulston},
  {Ryan}, {Rykoff}, {Sabater}, {Sakurikar}, {Salgado}, {Sanghi}, {Saunders},
  {Savchenko}, {Schwardt}, {Seifert-Eckert}, {Shih}, {Jain}, {Shukla}, {Sick},
  {Simpson}, {Singanamalla}, {Singer}, {Singhal}, {Sinha}, {Sip{\H{o}}cz},
  {Spitler}, {Stansby}, {Streicher}, {{\v{S}}umak}, {Swinbank}, {Taranu},
  {Tewary}, {Tremblay}, {Val-Borro}, {Van Kooten}, {Vasovi{\'c}}, {Verma}, {de
  Miranda Cardoso}, {Williams}, {Wilson}, {Winkel}, {Wood-Vasey}, {Xue},
  {Yoachim}, {Zhang}, {Zonca}, \& {Astropy Project Contributors}}]{astropy3}
{Astropy Collaboration}, {Price-Whelan}, A.~M., {Lim}, P.~L., {et~al.} 2022,
  \apj, 935, 167, \dodoi{10.3847/1538-4357/ac7c74}

\bibitem[{{Bellot Rubio} \& {Orozco Su{\'a}rez}(2012)}]{rubio2012}
{Bellot Rubio}, L.~R., \& {Orozco Su{\'a}rez}, D. 2012, \apj, 757, 19,
  \dodoi{10.1088/0004-637X/757/1/19}

\bibitem[{{Borrero} \& {Kobel}(2011)}]{Borrero2011}
{Borrero}, J.~M., \& {Kobel}, P. 2011, \aap, 527, A29,
  \dodoi{10.1051/0004-6361/201015634}

\bibitem[{Campbell(2023)}]{SIRE}
Campbell, R.~J. 2023, {SIR Explorer: a Python 3 visualisation tool for
  exploring SIR/DeSIRe inversions.}, 1.0,  Zenodo,
  \dodoi{10.5281/zenodo.7529086}

\bibitem[{{Campbell} {et~al.}(2021{\natexlab{a}}){Campbell}, {Mathioudakis},
  {Collados}, {Keys}, {Asensio Ramos}, {Nelson}, {Kuridze}, \&
  {Reid}}]{Campbell2021a}
{Campbell}, R.~J., {Mathioudakis}, M., {Collados}, M., {et~al.}
  2021{\natexlab{a}}, \aap, 647, A182, \dodoi{10.1051/0004-6361/202040028}

\bibitem[{{Campbell} {et~al.}(2021{\natexlab{b}}){Campbell}, {Shelyag},
  {Quintero Noda}, {Mathioudakis}, {Keys}, \& {Reid}}]{Campbell2021b}
{Campbell}, R.~J., {Shelyag}, S., {Quintero Noda}, C., {et~al.}
  2021{\natexlab{b}}, \aap, 654, A11, \dodoi{10.1051/0004-6361/202141421}

\bibitem[{{Collados}(1999)}]{crosstalk1999}
{Collados}, M. 1999, in Astronomical Society of the Pacific Conference Series,
  Vol. 184, Third Advances in Solar Physics Euroconference: Magnetic Fields and
  Oscillations, ed. B.~{Schmieder}, A.~{Hofmann}, \& J.~{Staude}, 3--22

\bibitem[{{Collados} {et~al.}(2012){Collados}, {L{\'o}pez}, {P{\'a}ez},
  {Hern{\'a}ndez}, {Reyes}, {Calcines}, {Ballesteros}, {D{\'\i}az}, {Denker},
  {Lagg}, {Schlichenmaier}, {Schmidt}, {Solanki}, {Strassmeier}, {von der
  L{\"u}he}, \& {Volkmer}}]{gregor2012}
{Collados}, M., {L{\'o}pez}, R., {P{\'a}ez}, E., {et~al.} 2012, Astronomische
  Nachrichten, 333, 872, \dodoi{10.1002/asna.201211738}

\bibitem[{{Danilovic} {et~al.}(2010){Danilovic}, {Beeck}, {Pietarila},
  {Sch{\"u}ssler}, {Solanki}, {Mart{\'\i}nez Pillet}, {Bonet}, {del Toro
  Iniesta}, {Domingo}, {Barthol}, {Berkefeld}, {Gandorfer}, {Kn{\"o}lker},
  {Schmidt}, \& {Title}}]{danilovic_imax_2010}
{Danilovic}, S., {Beeck}, B., {Pietarila}, A., {et~al.} 2010, \apjl, 723, L149,
  \dodoi{10.1088/2041-8205/723/2/L149}

\bibitem[{{Dominguez-Tagle} {et~al.}(2022){Dominguez-Tagle}, {Collados},
  {Lopez}, {Vaz Cedillo}, {Esteves}, {Grassin}, {Vega}, {Mato}, {Quintero},
  {Rodriguez}, {Regalado}, \& {Gonzalez}}]{DominguezTagle2022}
{Dominguez-Tagle}, C., {Collados}, M., {Lopez}, R., {et~al.} 2022, Journal of
  Astronomical Instrumentation, \dodoi{10.1142/s2251171722500143}

\bibitem[{{Gafeira} {et~al.}(2021){Gafeira}, {Orozco Su{\'a}rez}, {Mili{\'c}},
  {Quintero Noda}, {Ruiz Cobo}, \& {Uitenbroek}}]{ricardo2021}
{Gafeira}, R., {Orozco Su{\'a}rez}, D., {Mili{\'c}}, I., {et~al.} 2021, \aap,
  651, A31, \dodoi{10.1051/0004-6361/201936910}

\bibitem[{{Go{\v{s}}i{\'c}} {et~al.}(2022){Go{\v{s}}i{\'c}}, {Bellot Rubio},
  {Cheung}, {Orozco Su{\'a}rez}, {Katsukawa}, \& {del Toro
  Iniesta}}]{gosic2022}
{Go{\v{s}}i{\'c}}, M., {Bellot Rubio}, L.~R., {Cheung}, M.~C.~M., {et~al.}
  2022, \apj, 925, 188, \dodoi{10.3847/1538-4357/ac37be}

\bibitem[{{Go{\v{s}}i{\'c}} {et~al.}(2021){Go{\v{s}}i{\'c}}, {De Pontieu},
  {Bellot Rubio}, {Sainz Dalda}, \& {Pozuelo}}]{gosic2021}
{Go{\v{s}}i{\'c}}, M., {De Pontieu}, B., {Bellot Rubio}, L.~R., {Sainz Dalda},
  A., \& {Pozuelo}, S.~E. 2021, \apj, 911, 41, \dodoi{10.3847/1538-4357/abe7e0}

\bibitem[{{Harra} {et~al.}(2010){Harra}, {Magara}, {Hara}, {Tsuneta},
  {Okamoto}, \& {Wallace}}]{harra2010}
{Harra}, L.~K., {Magara}, T., {Hara}, H., {et~al.} 2010, \solphys, 263, 105,
  \dodoi{10.1007/s11207-010-9548-x}

\bibitem[{Harris {et~al.}(2020)Harris, Millman, van~der Walt, Gommers,
  Virtanen, Cournapeau, Wieser, Taylor, Berg, Smith, Kern, Picus, Hoyer, van
  Kerkwijk, Brett, Haldane, del R{\'{i}}o, Wiebe, Peterson,
  G{\'{e}}rard-Marchant, Sheppard, Reddy, Weckesser, Abbasi, Gohlke, \&
  Oliphant}]{numpy}
Harris, C.~R., Millman, K.~J., van~der Walt, S.~J., {et~al.} 2020, Nature, 585,
  357, \dodoi{10.1038/s41586-020-2649-2}

\bibitem[{Hunter(2007)}]{mpl}
Hunter, J.~D. 2007, Computing in Science and Engineering, 9, 90,
  \dodoi{10.1109/MCSE.2007.55}

\bibitem[{{Khomenko} {et~al.}(2003){Khomenko}, {Collados}, {Solanki}, {Lagg},
  \& {Trujillo Bueno}}]{khomenko2003}
{Khomenko}, E.~V., {Collados}, M., {Solanki}, S.~K., {Lagg}, A., \& {Trujillo
  Bueno}, J. 2003, \aap, 408, 1115, \dodoi{10.1051/0004-6361:20030604}

\bibitem[{{Kianfar} {et~al.}(2018){Kianfar}, {Jafarzadeh}, {Mirtorabi}, \&
  {Riethm{\"u}ller}}]{kianfar2018}
{Kianfar}, S., {Jafarzadeh}, S., {Mirtorabi}, M.~T., \& {Riethm{\"u}ller},
  T.~L. 2018, \solphys, 293, 123, \dodoi{10.1007/s11207-018-1341-2}

\bibitem[{{Kleint} {et~al.}(2020){Kleint}, {Berkefeld}, {Esteves}, {Sonner},
  {Volkmer}, {Gerber}, {Kr{\"a}mer}, {Grassin}, \& {Berdyugina}}]{kleint2020}
{Kleint}, L., {Berkefeld}, T., {Esteves}, M., {et~al.} 2020, \aap, 641, A27,
  \dodoi{10.1051/0004-6361/202038208}

\bibitem[{{Lagg} {et~al.}(2016){Lagg}, {Solanki}, {Doerr}, {Mart{\'\i}nez
  Gonz{\'a}lez}, {Riethm{\"u}ller}, {Collados Vera}, {Schlichenmaier}, {Orozco
  Su{\'a}rez}, {Franz}, {Feller}, {Kuckein}, {Schmidt}, {Asensio Ramos},
  {Pastor Yabar}, {von der L{\"u}he}, {Denker}, {Balthasar}, {Volkmer},
  {Staude}, {Hofmann}, {Strassmeier}, {Kneer}, {Waldmann}, {Borrero},
  {Sobotka}, {Verma}, {Louis}, {Rezaei}, {Soltau}, {Berkefeld}, {Sigwarth},
  {Schmidt}, {Kiess}, \& {Nicklas}}]{lagg2016}
{Lagg}, A., {Solanki}, S.~K., {Doerr}, H.~P., {et~al.} 2016, \aap, 596, A6,
  \dodoi{10.1051/0004-6361/201628489}

\bibitem[{{Ledvina} {et~al.}(2022){Ledvina}, {Kazachenko}, {Criscuoli},
  {Tilipman}, {Ermolli}, {Falco}, {Guglielmino}, {Jafarzadeh}, {van der Voort},
  \& {Zuccarello}}]{ledvina2022}
{Ledvina}, V.~E., {Kazachenko}, M.~D., {Criscuoli}, S., {et~al.} 2022, \apj,
  934, 38, \dodoi{10.3847/1538-4357/ac7785}

\bibitem[{{Lites} {et~al.}(2008){Lites}, {Kubo}, {Socas-Navarro}, {Berger},
  {Frank}, {Shine}, {Tarbell}, {Title}, {Ichimoto}, {Katsukawa}, {Tsuneta},
  {Suematsu}, {Shimizu}, \& {Nagata}}]{lites2008}
{Lites}, B.~W., {Kubo}, M., {Socas-Navarro}, H., {et~al.} 2008, \apj, 672,
  1237, \dodoi{10.1086/522922}

\bibitem[{{Livingston} \& {Wallace}(1991)}]{atlas}
{Livingston}, W., \& {Wallace}, L. 1991, {An atlas of the solar spectrum in the
  infrared from 1850 to 9000 cm-1 (1.1 to 5.4 micrometer)}

\bibitem[{{Mart{\'\i}nez Gonz{\'a}lez} {et~al.}(2008){Mart{\'\i}nez
  Gonz{\'a}lez}, {Asensio Ramos}, {L{\'o}pez Ariste}, \& {Manso
  Sainz}}]{marian2008_mu}
{Mart{\'\i}nez Gonz{\'a}lez}, M.~J., {Asensio Ramos}, A., {L{\'o}pez Ariste},
  A., \& {Manso Sainz}, R. 2008, \aap, 479, 229,
  \dodoi{10.1051/0004-6361:20078500}

\bibitem[{{Mart{\'\i}nez Gonz{\'a}lez} {et~al.}(2012){Mart{\'\i}nez
  Gonz{\'a}lez}, {Manso Sainz}, {Asensio Ramos}, \& {Hijano}}]{marian2012}
{Mart{\'\i}nez Gonz{\'a}lez}, M.~J., {Manso Sainz}, R., {Asensio Ramos}, A., \&
  {Hijano}, E. 2012, \apj, 755, 175, \dodoi{10.1088/0004-637X/755/2/175}

\bibitem[{{Mart{\'\i}nez Gonz{\'a}lez} {et~al.}(2016){Mart{\'\i}nez
  Gonz{\'a}lez}, {Pastor Yabar}, {Lagg}, {Asensio Ramos}, {Collados},
  {Solanki}, {Balthasar}, {Berkefeld}, {Denker}, {Doerr}, {Feller}, {Franz},
  {Gonz{\'a}lez Manrique}, {Hofmann}, {Kneer}, {Kuckein}, {Louis}, {von der
  L{\"u}he}, {Nicklas}, {Orozco}, {Rezaei}, {Schlichenmaier}, {Schmidt},
  {Schmidt}, {Sigwarth}, {Sobotka}, {Soltau}, {Staude}, {Strassmeier}, {Verma},
  {Waldman}, \& {Volkmer}}]{marian2016}
{Mart{\'\i}nez Gonz{\'a}lez}, M.~J., {Pastor Yabar}, A., {Lagg}, A., {et~al.}
  2016, \aap, 596, A5, \dodoi{10.1051/0004-6361/201628449}

\bibitem[{{Nelson} {et~al.}(2021){Nelson}, {Campbell}, \&
  {Mathioudakis}}]{Nelson2021}
{Nelson}, C.~J., {Campbell}, R.~J., \& {Mathioudakis}, M. 2021, \aap, 654, A50,
  \dodoi{10.1051/0004-6361/202141368}

\bibitem[{{Orozco Su{\'a}rez} \& {Bellot Rubio}(2012)}]{suarez2012}
{Orozco Su{\'a}rez}, D., \& {Bellot Rubio}, L.~R. 2012, \apj, 751, 2,
  \dodoi{10.1088/0004-637X/751/1/2}

\bibitem[{{Quintero Noda} {et~al.}(2021){Quintero Noda}, {Barklem}, {Gafeira},
  {Ruiz Cobo}, {Collados}, {Carlsson}, {Mart{\'\i}nez Pillet}, {Orozco
  Su{\'a}rez}, {Uitenbroek}, \& {Katsukawa}}]{quintero2021}
{Quintero Noda}, C., {Barklem}, P.~S., {Gafeira}, R., {et~al.} 2021, \aap, 652,
  A161, \dodoi{10.1051/0004-6361/202037735}

\bibitem[{{Quintero Noda} {et~al.}(2022){Quintero Noda}, {Schlichenmaier},
  {Bellot Rubio}, {L{\"o}fdahl}, {Khomenko}, {Jur{\v{c}}{\'a}k}, {Leenaarts},
  {Kuckein}, {Gonz{\'a}lez Manrique}, {Gun{\'a}r}, {Nelson}, {de la Cruz
  Rodr{\'\i}guez}, {Tziotziou}, {Tsiropoula}, {Aulanier}, {Aboudarham},
  {Allegri}, {Alsina Ballester}, {Amans}, {Asensio Ramos}, {Bail{\'e}n},
  {Balaguer}, {Baldini}, {Balthasar}, {Barata}, {Barczynski}, {Barreto
  Cabrera}, {Baur}, {B{\'e}chet}, {Beck}, {Bel{\'\i}o-As{\'\i}n},
  {Bello-Gonz{\'a}lez}, {Belluzzi}, {Bentley}, {Berdyugina}, {Berghmans},
  {Berlicki}, {Berrilli}, {Berkefeld}, {Bettonvil}, {Bianda}, {Bienes
  P{\'e}rez}, {Bonaque-Gonz{\'a}lez}, {Braj{\v{s}}a}, {Bommier}, {Bourdin},
  {Burgos Mart{\'\i}n}, {Calchetti}, {Calcines}, {Calvo Tovar}, {Campbell},
  {Carballo-Mart{\'\i}n}, {Carbone}, {Carlin}, {Carlsson}, {Castro L{\'o}pez},
  {Cavaller}, {Cavallini}, {Cauzzi}, {Cecconi}, {Chulani}, {Cirami},
  {Consolini}, {Coretti}, {Cosentino}, {C{\'o}zar-Castellano}, {Dalmasse},
  {Danilovic}, {De Juan Ovelar}, {Del Moro}, {del Pino Alem{\'a}n}, {del Toro
  Iniesta}, {Denker}, {Dhara}, {Di Marcantonio}, {D{\'\i}az Baso}, {Diercke},
  {Dineva}, {D{\'\i}az-Garc{\'\i}a}, {Doerr}, {Doyle}, {Erdelyi}, {Ermolli},
  {Escobar Rodr{\'\i}guez}, {Esteban Pozuelo}, {Faurobert}, {Felipe}, {Feller},
  {Feijoo Amoedo}, {Femen{\'\i}a Castell{\'a}}, {Fernandes}, {Ferro
  Rodr{\'\i}guez}, {Figueroa}, {Fletcher}, {Franco Ordovas}, {Gafeira},
  {Gardenghi}, {Gelly}, {Giorgi}, {Gisler}, {Giovannelli}, {Gonz{\'a}lez},
  {Gonz{\'a}lez}, {Gonz{\'a}lez-Cava}, {Gonz{\'a}lez Garc{\'\i}a},
  {G{\"o}m{\"o}ry}, {Gracia}, {Grauf}, {Greco}, {Grivel}, {Guerreiro},
  {Guglielmino}, {Hammerschlag}, {Hanslmeier}, {Hansteen}, {Heinzel},
  {Hern{\'a}ndez-Delgado}, {Hern{\'a}ndez Su{\'a}rez}, {Hidalgo}, {Hill},
  {Hizberger}, {Hofmeister}, {J{\"a}gers}, {Janett}, {Jarolim}, {Jess},
  {Jim{\'e}nez Mej{\'\i}as}, {Jolissaint}, {Kamlah}, {Kapit{\'a}n},
  {Ka{\v{s}}parov{\'a}}, {Keller}, {Kentischer}, {Kiselman}, {Kleint},
  {Klvana}, {Kontogiannis}, {Krishnappa}, {Ku{\v{c}}era}, {Labrosse}, {Lagg},
  {Landi Degl'Innocenti}, {Langlois}, {Lafon}, {Laforgue}, {Le Men}, {Lepori},
  {Lepreti}, {Lindberg}, {Lilje}, {L{\'o}pez Ariste}, {L{\'o}pez
  Fern{\'a}ndez}, {L{\'o}pez Jim{\'e}nez}, {L{\'o}pez L{\'o}pez}, {Manso
  Sainz}, {Marassi}, {Marco de la Rosa}, {Marino}, {Marrero}, {Mart{\'\i}n},
  {Mart{\'\i}n G{\'a}lvez}, {Mart{\'\i}n Hernando}, {Masciadri}, {Mart{\'\i}nez
  Gonz{\'a}lez}, {Matta-G{\'o}mez}, {Mato}, {Mathioudakis}, {Matthews}, {Mein},
  {Merlos Garc{\'\i}a}, {Moity}, {Montilla}, {Molinaro}, {Molodij}, {Montoya},
  {Munari}, {Murabito}, {N{\'u}{\~n}ez Cagigal}, {Oliviero}, {Orozco
  Su{\'a}rez}, {Ortiz}, {Padilla-Hern{\'a}ndez}, {Pa{\'e}z Ma{\~n}{\'a}},
  {Paletou}, {Pancorbo}, {Pastor Ca{\~n}edo}, {Pastor Yabar}, {Peat},
  {Pedichini}, {Peixinho}, {Pe{\~n}ate}, {P{\'e}rez de Taoro}, {Peter},
  {Petrovay}, {Piazzesi}, {Pietropaolo}, {Pleier}, {Poedts}, {P{\"o}tzi},
  {Podladchikova}, {Prieto}, {Quintero Nehrkorn}, {Ramelli}, {Ramos Sapena},
  {Rasilla}, {Reardon}, {Rebolo}, {Regalado Olivares}, {Reyes
  Garc{\'\i}a-Talavera}, {Riethm{\"u}ller}, {Rimmele}, {Rodr{\'\i}guez
  Delgado}, {Rodr{\'\i}guez Gonz{\'a}lez}, {Rodr{\'\i}guez-Losada},
  {Rodr{\'\i}guez Ramos}, {Romano}, {Roth}, {Rouppe van der Voort}, {Rudawy},
  {Ruiz de Galarreta}, {Ryb{\'a}k}, {Salvade}, {S{\'a}nchez-Capuchino},
  {S{\'a}nchez Rodr{\'\i}guez}, {Sangiorgi}, {Say{\`e}de}, {Scharmer},
  {Scheiffelen}, {Schmidt}, {Schmieder}, {Scir{\`e}}, {Scuderi}, {Siegel},
  {Sigwarth}, {Sim{\~o}es}, {Snik}, {Sliepen}, {Sobotka}, {Socas-Navarro},
  {Sola La Serna}, {Solanki}, {Soler Trujillo}, {Soltau}, {Sordini}, {Sosa
  M{\'e}ndez}, {Stangalini}, {Steiner}, {Stenflo}, {{\v{S}}t{\v{e}}p{\'a}n},
  {Strassmeier}, {Sudar}, {Suematsu}, {S{\"u}tterlin}, {Tallon}, {Temmer},
  {Tenegi}, {Tritschler}, {Trujillo Bueno}, {Turchi}, {Utz}, {van Harten}, {van
  Noort}, {van Werkhoven}, {Vansintjan}, {Vaz Cedillo}, {Vega Reyes}, {Verma},
  {Veronig}, {Viavattene}, {Vitas}, {V{\"o}gler}, {von der L{\"u}he},
  {Volkmer}, {Waldmann}, {Walton}, {Wisniewska}, {Zeman}, {Zeuner}, {Zhang},
  {Zuccarello}, \& {Collados}}]{EST}
{Quintero Noda}, C., {Schlichenmaier}, R., {Bellot Rubio}, L.~R., {et~al.}
  2022, \aap, 666, A21, \dodoi{10.1051/0004-6361/202243867}

\bibitem[{{Rast} {et~al.}(2021){Rast}, {Bello Gonz{\'a}lez}, {Bellot Rubio},
  {Cao}, {Cauzzi}, {Deluca}, {de Pontieu}, {Fletcher}, {Gibson}, {Judge},
  {Katsukawa}, {Kazachenko}, {Khomenko}, {Landi}, {Mart{\'\i}nez Pillet},
  {Petrie}, {Qiu}, {Rachmeler}, {Rempel}, {Schmidt}, {Scullion}, {Sun},
  {Welsch}, {Andretta}, {Antolin}, {Ayres}, {Balasubramaniam}, {Ballai},
  {Berger}, {Bradshaw}, {Campbell}, {Carlsson}, {Casini}, {Centeno}, {Cranmer},
  {Criscuoli}, {Deforest}, {Deng}, {Erd{\'e}lyi}, {Fedun}, {Fischer},
  {Gonz{\'a}lez Manrique}, {Hahn}, {Harra}, {Henriques}, {Hurlburt}, {Jaeggli},
  {Jafarzadeh}, {Jain}, {Jefferies}, {Keys}, {Kowalski}, {Kuckein}, {Kuhn},
  {Kuridze}, {Liu}, {Liu}, {Longcope}, {Mathioudakis}, {McAteer}, {McIntosh},
  {McKenzie}, {Miralles}, {Morton}, {Muglach}, {Nelson}, {Panesar}, {Parenti},
  {Parnell}, {Poduval}, {Reardon}, {Reep}, {Schad}, {Schmit}, {Sharma},
  {Socas-Navarro}, {Srivastava}, {Sterling}, {Suematsu}, {Tarr}, {Tiwari},
  {Tritschler}, {Verth}, {Vourlidas}, {Wang}, {Wang}, {NSO and DKIST Project},
  {DKIST Instrument Scientists}, {DKIST Science Working Group}, \& {DKIST
  Critical Science Plan Community}}]{dkistcsp}
{Rast}, M.~P., {Bello Gonz{\'a}lez}, N., {Bellot Rubio}, L., {et~al.} 2021,
  \solphys, 296, 70, \dodoi{10.1007/s11207-021-01789-2}

\bibitem[{{Rimmele} {et~al.}(2020){Rimmele}, {Warner}, {Keil}, {Goode},
  {Kn{\"o}lker}, {Kuhn}, {Rosner}, {McMullin}, {Casini}, {Lin}, {W{\"o}ger},
  {von der L{\"u}he}, {Tritschler}, {Davey}, {de Wijn}, {Elmore}, {Fehlmann},
  {Harrington}, {Jaeggli}, {Rast}, {Schad}, {Schmidt}, {Mathioudakis},
  {Mickey}, {Anan}, {Beck}, {Marshall}, {Jeffers}, {Oschmann}, {Beard},
  {Berst}, {Cowan}, {Craig}, {Cross}, {Cummings}, {Donnelly}, {de Vanssay},
  {Eigenbrot}, {Ferayorni}, {Foster}, {Galapon}, {Gedrites}, {Gonzales},
  {Goodrich}, {Gregory}, {Guzman}, {Guzzo}, {Hegwer}, {Hubbard}, {Hubbard},
  {Johansson}, {Johnson}, {Liang}, {Liang}, {McQuillen}, {Mayer}, {Newman},
  {Onodera}, {Phelps}, {Puentes}, {Richards}, {Rimmele}, {Sekulic}, {Shimko},
  {Simison}, {Smith}, {Starman}, {Sueoka}, {Summers}, {Szabo}, {Szabo},
  {Wampler}, {Williams}, \& {White}}]{DKIST}
{Rimmele}, T.~R., {Warner}, M., {Keil}, S.~L., {et~al.} 2020, \solphys, 295,
  172, \dodoi{10.1007/s11207-020-01736-7}

\bibitem[{{Ruiz Cobo} \& {del Toro Iniesta}(1992)}]{SIR}
{Ruiz Cobo}, B., \& {del Toro Iniesta}, J.~C. 1992, \apj, 398, 375,
  \dodoi{10.1086/171862}

\bibitem[{{Ruiz Cobo} {et~al.}(2022){Ruiz Cobo}, {Quintero Noda}, {Gafeira},
  {Uitenbroek}, {Orozco Su{\'a}rez}, \& {P{\'a}ez Ma{\~n}{\'a}}}]{desire}
{Ruiz Cobo}, B., {Quintero Noda}, C., {Gafeira}, R., {et~al.} 2022, \aap, 660,
  A37, \dodoi{10.1051/0004-6361/202140877}

\bibitem[{{S{\'a}nchez Almeida} \& {Mart{\'\i}nez
  Gonz{\'a}lez}(2011)}]{almeida2011}
{S{\'a}nchez Almeida}, J., \& {Mart{\'\i}nez Gonz{\'a}lez}, M. 2011, in
  Astronomical Society of the Pacific Conference Series, Vol. 437, Solar
  Polarization 6, ed. J.~R. {Kuhn}, D.~M. {Harrington}, H.~{Lin}, S.~V.
  {Berdyugina}, J.~{Trujillo-Bueno}, S.~L. {Keil}, \& T.~{Rimmele}, 451.
\newblock \doarXiv{1105.0387}

\bibitem[{{Schmidt} {et~al.}(2012){Schmidt}, {von der L{\"u}he}, {Volkmer},
  {Denker}, {Solanki}, {Balthasar}, {Bello Gonzalez}, {Berkefeld}, {Collados},
  {Fischer}, {Halbgewachs}, {Heidecke}, {Hofmann}, {Kneer}, {Lagg}, {Nicklas},
  {Popow}, {Puschmann}, {Schmidt}, {Sigwarth}, {Sobotka}, {Soltau}, {Staude},
  {Strassmeier}, \& {Waldmann }}]{Schmidt2012}
{Schmidt}, W., {von der L{\"u}he}, O., {Volkmer}, R., {et~al.} 2012,
  Astronomische Nachrichten, 333, 796, \dodoi{10.1002/asna.201211725}

\bibitem[{{Steiner} \& {Rezaei}(2012)}]{steiner2012}
{Steiner}, O., \& {Rezaei}, R. 2012, in Astronomical Society of the Pacific
  Conference Series, Vol. 456, Fifth Hinode Science Meeting, ed. L.~{Golub},
  I.~{De Moortel}, \& T.~{Shimizu}, 3.
\newblock \doarXiv{1202.4040}

\bibitem[{{Steiner} {et~al.}(2008){Steiner}, {Rezaei}, {Schaffenberger}, \&
  {Wedemeyer-B{\"o}hm}}]{steiner2008}
{Steiner}, O., {Rezaei}, R., {Schaffenberger}, W., \& {Wedemeyer-B{\"o}hm}, S.
  2008, \apjl, 680, L85, \dodoi{10.1086/589740}

\bibitem[{{Trelles Arjona} {et~al.}(2021){Trelles Arjona}, {Ruiz Cobo}, \&
  {Mart{\'\i}nez Gonz{\'a}lez}}]{lines2021}
{Trelles Arjona}, J.~C., {Ruiz Cobo}, B., \& {Mart{\'\i}nez Gonz{\'a}lez},
  M.~J. 2021, \aap, 648, A68, \dodoi{10.1051/0004-6361/202038941}

\bibitem[{{V{\"o}gler} {et~al.}(2005){V{\"o}gler}, {Shelyag}, {Sch{\"u}ssler},
  {Cattaneo}, {Emonet}, \& {Linde}}]{vogler2005}
{V{\"o}gler}, A., {Shelyag}, S., {Sch{\"u}ssler}, M., {et~al.} 2005, \aap, 429,
  335, \dodoi{10.1051/0004-6361:20041507}

\end{thebibliography}
\bibliographystyle{aasjournal}

\end{document}